\documentclass[aps,pre,10pt,twocolumn,nofootinbib]{revtex4-2}

\usepackage{amsmath,amssymb,times,hyperref,graphicx,tabularray,cancel}
\hypersetup{colorlinks=true, linkcolor=blue, citecolor=blue, filecolor=blue, urlcolor=blue}

\newcommand{\und}[1]{\underline{#1}}

\newcommand{\eg}{\textit{e.g.}, }
\newcommand{\ie}{\textit{i.e.}, }
\newcommand{\cf}{\textit{cf.} }

\renewcommand{\b}{\mathrm{b}}
\renewcommand{\d}{\mathrm{d}}
\newcommand{\f}{\mathrm{f}}
\renewcommand{\i}{\mathrm{i}}
\newcommand{\p}{\mathrm{p}}
\newcommand{\T}{\mathrm{T}}

\newcommand{\kt}{k_\mathrm{B}\mathbb{T}}
\newcommand{\sgn}{\,\mathrm{sgn}\,}

\newcommand{\vecb}{\mathbf{b}}
\newcommand{\vecB}{\mathbf{B}}
\newcommand{\vecF}{\mathbf{F}}

\newcommand{\vecf}{\mathbf{f}}
\newcommand{\vecg}{\mathbf{g}}
\newcommand{\vecI}{\mathbf{I}}
\newcommand{\vecj}{\mathbf{j}}
\newcommand{\veck}{\mathbf{k}}
\newcommand{\vecm}{\mathbf{m}}
\newcommand{\vecM}{\mathbf{M}}
\newcommand{\vecn}{\mathbf{n}}
\newcommand{\vecO}{\mathbf{O}}
\newcommand{\vecR}{\mathbf{R}}
\newcommand{\vecr}{\mathbf{r}}
\newcommand{\vecQ}{\mathbf{Q}}
\newcommand{\vecU}{\mathbf{U}}
\newcommand{\vecu}{\mathbf{u}}
\newcommand{\vecV}{\mathbf{V}}
\newcommand{\vecv}{\mathbf{v}}

\newcommand{\vecy}{\mathbf{y}}

\newcommand{\bbeta}{\boldsymbol{\beta}}
\newcommand{\bgamma}{\boldsymbol{\gamma}}
\newcommand{\bGamma}{\boldsymbol{\Gamma}}
\newcommand{\bDelta}{\boldsymbol{\Delta}}
\newcommand{\bsigma}{\boldsymbol{\sigma}}
\newcommand{\bxi}{\boldsymbol{\xi}}
\newcommand{\bXi}{\boldsymbol{\Xi}}
\newcommand{\bta}{\boldsymbol{\eta}}
\newcommand{\bTa}{\mathbf{H}}
\newcommand{\btau}{\boldsymbol{\tau}}
\newcommand{\bomega}{\boldsymbol{\omega}}
\newcommand{\bnu}{\boldsymbol{\nu}}

\newcommand{\calO}{\mathcal{O}}
\newcommand{\grad}{\boldsymbol{\nabla}}

\begin{document} 

\title{Brownian motion with geometry-dependent hydrodynamic memory} 

\author{Benjamin Sorkin}
\thanks{B.S. and G.T. contributed equally}
\email{bs4171@princeton.edu}
\affiliation{Princeton Center for Theoretical Science, Princeton University, Princeton, NJ 08544, USA}

\author{G\"unther Turk}
\thanks{B.S. and G.T. contributed equally}
\email{guenther.turk@princeton.edu}
\affiliation{Princeton Materials Institute, Princeton University, Princeton, NJ 08544, USA}

\author{Howard A. Stone}
\email{hastone@princeton.edu}
\affiliation{Department of Mechanical and Aerospace Engineering, Princeton University, Princeton, NJ 08544, USA}

\begin{abstract}
    Micron-sized particles moving through a fluid are subject to viscous resistance and thermal fluctuations. Beyond steady Stokes friction, colloids also exhibit hydrodynamic memory effects arising from conservation laws of the surrounding fluid. Although fluid-flow problems in complex geometries and confinements are often highly involved, numerical or approximate solutions can be obtained; translating these solutions into closed-form equations of motion for individual colloids, however, is rarely possible. Here, we develop a theoretical framework that provides both underdamped and overdamped colloidal dynamics from the solution of an arbitrary flow problem: (i) Using the Lorentz reciprocal theorem, we express the colloidal equation of motion in terms of the geometry's Green's function and the fluid-flow profile. This formulation yields both the corresponding Stokes drag as well as a Basset-like hydrodynamic memory contribution. (ii) As colloid inertia is often negligible, we furthermore derive the corresponding overdamped (colloid-inertia-less) limit, which inherits the Stokes and Basset-like resistances and additionally gives rise to a spurious drift. We propose applications of this framework to passive and active particles subject to involved confinements and problem geometries. 
\end{abstract}

\maketitle

\section{Introduction}

A pervasive model for diffusion is Langevin dynamics~\cite{SchussBOOK2010}, a stochastic differential equation for the fluctuating microscopic system configurations under noise. It appears in diverse modeling and numerical contexts, ranging from quantitative finance~\cite{OksendalBOOK2003,LaxBOOK2006} and population dynamics~\cite{TsimringPRL1996,HallatschekPNAS2011,MarchiPNAS2021} to generative models~\cite{SohlDickesteinICML15,BiroliJSTAT2023} and computation~\cite{AiferNPJ2024}. In modeling certain physical systems, it is possible to derive the Langevin equation on rigorous grounds starting from the system's microscopic details, using methods such as the the Mori-Zwanzig projection formalism~\cite{ZwanzigBOOK2001}. For a colloid immersed in a fluid, the particle's equation of motion can be obtained by integrating the viscous stress over its surface~\cite{HaugeJSP1973}. This requires detailed knowledge of the flow field around the particle, which is often accessible only computationally. While numerical integration of the stress yields the particle dynamics, it obscures the physical interpretation of the hydrodynamic effects at play and how they modify the conventional Langevin equation. How can we derive a physically transparent Langevin equation that captures hydrodynamic effects across diverse fluid geometries, from nanocolloids under tight confinement~\cite{BocquetCSR2010} to bacteria swimming within hydration layers~\cite{BeerME2019}?

In devising such a procedure, one should take note of fluid conservation laws, as they introduce important subtleties into colloidal dynamics. Consider a colloid in an unbounded fluid, fully described by the fluctuating Basset-Boussinesq-Oseen equation (FBBOE)~\cite{HaugeJSP1973,ClercxPRA1992}. In addition to the usual (steady) Stokes drag of the conventional Langevin equation, the viscous friction in the FBBOE contains an added-mass contribution from the inertia of the displaced fluid and a Basset memory term arising from the diffusion of vorticity. While the former effectively amounts to increasing the colloid's mass, the latter gives rise to long-lived, power-law hydrodynamic memory. This memory manifests in experiments through colloidal velocity autocorrelation~\cite{PaulJPA1981} and diffusivity~\cite{LukicPRL2005} decaying as a power law in time. Amid these complications, it is widely recognized that the inertia of the colloid is negligible once a colloid diffuses a distance of its own size~\cite{WeitzPRL1989}. As the colloidal inertia is (justifiably) routinely eliminated, what is the fate of the power-law hydrodynamic memory in the overdamped (colloid-inertia-less) limit?

\begin{figure}
    \centering
    \includegraphics[width=0.99\linewidth]{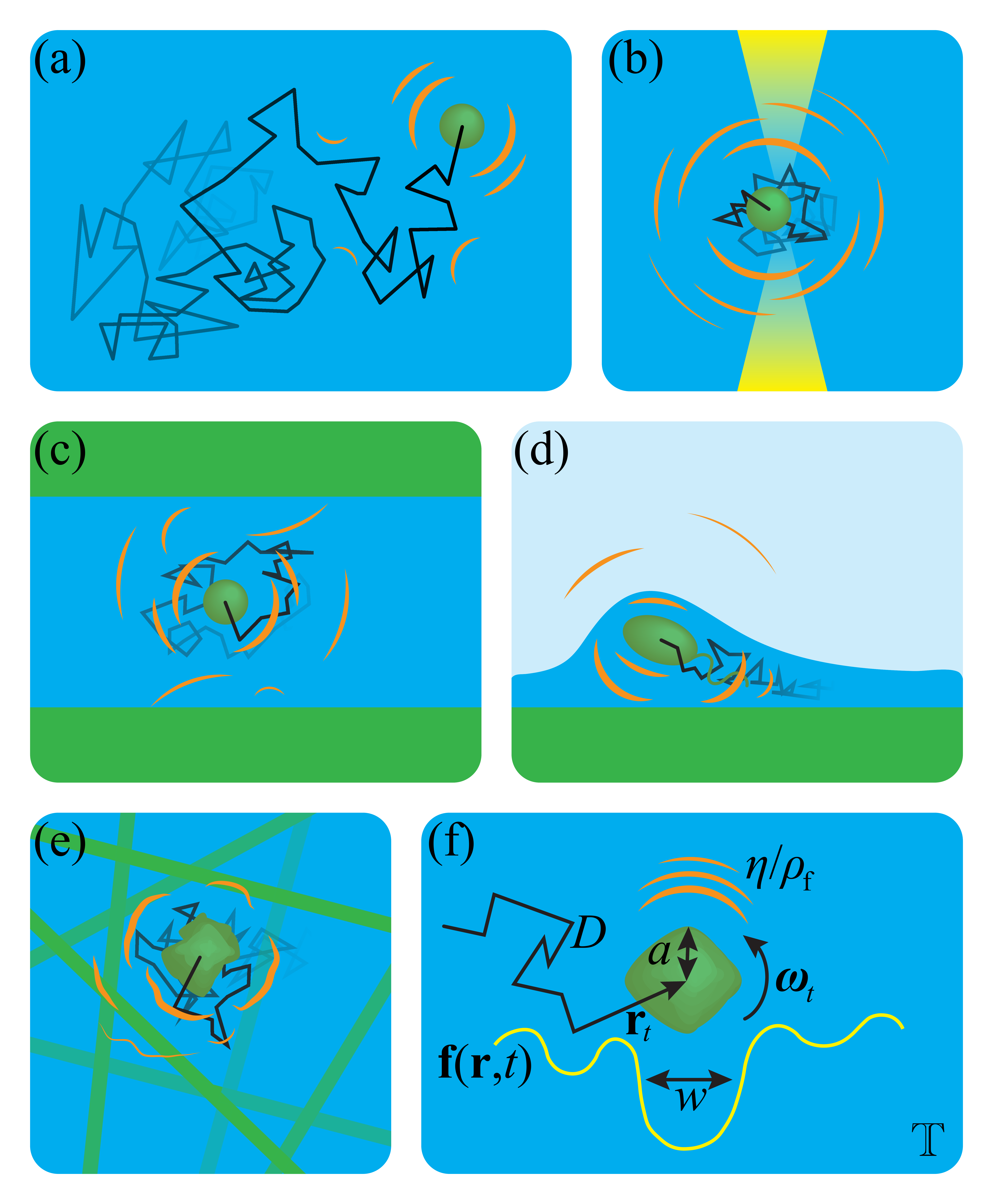}
    \caption{Illustration of systems to which we envision an application of our formalism. (a) Freely diffusing particle in an infinite fluid. (b) An optically trapped particle in an infinite fluid. (c) Colloid within a confined fluid. (d) A nonspherical active particle subject to wetting. (e) A nonspherical particle diffusing in a viscoelastic medium. We are especially interested in the diffusion of vorticity through the fluid (indicated by orange arcs), which must be compared to the particle diffusion (illustrated trajectories) to assess the importance of Basset memory. (f) Key system parameters: We investigate the particle time-dependent positional $\vecr_t$ and solid-angle $\bomega_t$ dynamics. The colloid is subject to external force $\vecf(\vecr,t)$ whose typical lengthscale is $w$. The particle's linear size is $a$, and the fluids temperature, viscosity, and density are $\kt$, $\eta$, and $\rho_\f$, respectively. The dimensionless parameter that will allow us to obtain the overdamped limit encodes the particle diffusivity $D\sim(\eta a)^{-1}$, the fluid's momentum diffusivity (kinematic viscosity) $\eta/\rho_\f$, particle size $a$, and trap halfwidth $w$.}
    \label{fig:illust}
\end{figure}

In this work, we provide a theoretical framework for tackling both questions. Using the Lorentz reciprocal theorem (RT)~\cite{TurkJFM2025}, we derive the full Langevin equation of a particle in an arbitrary flow (Eq.~\eqref{eq:RT-dimensional}). It explicitly contains the steady-Stokes friction of the particle as well as a hydrodynamic memory kernel in terms of the geometry's Green's function and fluid flow profile. Upon nondimensionalization, we identify the inverse Schmidt number\,---\,the ratio of colloid and fluid-momentum diffusivities (Eq.~\eqref{eq:epsilon}), underlying the crossover to the colloid-inertia-less limit\,---\,as the small parameter controlling the magnitude of hydrodynamic memory. We then employ stochastic Taylor expansions~\cite{KloedenBOOK1992,SchussBOOK2010,GardinerBOOK2009} in small inverse Schmidt number to derive the overdamped (colloid-inertia-less) Langevin equation for the given geometry (Eq.~\eqref{eq:RT_overdamped_unit}). In addition to the Stokes flow, linear response of velocity to force and noise, we obtain the corrections from hydrodynamic memory of past applied force and noise, as well as the corresponding spurious drift~\cite{LauPRE2007}. For clarity, we summarize our results along with relevant known formulae in Table~\ref{tab:summary}, and provide an illustration of systems we believe this formalism may be applicable to in Fig.~\ref{fig:illust}.

\begin{table*}
    \centering
    \begin{tblr}{|c|c|c|c|}
        \hline
            U1 & Langevin equation& $\frac23\kappa\epsilon\frac{\d\vecV_T}{\d T}=\vecF-\vecV_T + \mathrm{noise}$ & \cite{SchussBOOK2010,PathriaBOOK2011}\\
        \hline
            U2 & Free particle ($\vecF=\mathbf0$) & $\langle\vecV_T\vecV_0\rangle= \frac3{2\kappa\epsilon}e^{-3T/(2\kappa\epsilon)}\vecI$ &\cite{SchussBOOK2010,PathriaBOOK2011} \\
        \hline
        \hline
            U3 & FBBOE & $\kappa\epsilon\frac{\d\vecV_T}{\d T}=\vecF-\vecV_T-\epsilon^{1/2}\int_{-\infty}^T\frac{\d \vecV_T'}{\sqrt{T-T'}} +\mathrm{noise}$ & \cite{LandauBOOK1987,ClercxPRA1992}\\
        \hline
            U4 & Free particle ($\vecF=\mathbf0$) & $\langle\vecV_T\vecV_0\rangle\sim \left[\cancelto{}{\frac1{\kappa\epsilon}e^{-T/(\kappa\epsilon)}}+\frac{\epsilon^{1/2}}{2T^{3/2}}\right]\vecI$ & \cite{AdlerPRL1967,AdlerPRA1970,ErnstPRL1970,HinchJFM1975,BoonPLA1976,BouillerJPF1978,PaulJPA1981}\\
        \hline\hline
            U5 & RT-Langevin$*$ & $\kappa\epsilon\frac{\d\vecV_T}{\d T}=\vecF-\bGamma(\vecR_T)\cdot\vecV_T-\epsilon^{1/2}\mathcal{A}\vecV_{T} +\mathrm{noise}$ & Eqs.~\eqref{eq:RT-dimensional}, \eqref{eq:RT-Langevin}\\
        \hline\hline\hline
        O1 & Langevin equation & $\frac {\d \vecR_T}{\d T}=\vecF + \mathrm{noise}$ & \cite{SchussBOOK2010,OksendalBOOK2003,GardinerBOOK2009}\\
        \hline
            O2 & Trapped particle ($\vecF=-\vecR$) & $\langle\vecR_T\vecR_0\rangle= e^{-T}\vecI$ & \cite{SchussBOOK2010,OksendalBOOK2003} \\
        \hline
        \hline
            O3 & OBLE$\dagger$ & $\frac {\d \vecR_T}{\d T}=\vecF-\epsilon^{1/2}\int_{-\infty}^T\frac{ \d\vecF(\vecR_{T'})}{\sqrt{T-T'}}+ \mathrm{noise}$ & \cite{LETTER} and Eq.~\eqref{eq:OBLE}\\
        \hline
            O4 & Trapped particle$\dagger$ ($\vecF=-\vecR$) & $\langle\vecR_T\vecR_0\rangle\sim \left[e^{-T}-\frac{\epsilon^{1/2}}{2T^{3/2}}\right]\vecI$ & \cite{LETTER,FranoschNATURE2011} and Eq.~\eqref{eq:RR_trap} \\
        \hline\hline
            O5 & RT-Langevin$*$ & $\frac{\d\vecR_T}{\d T}=\bGamma^{-1}\cdot\vecF+\grad\cdot\bGamma^{-1}-\epsilon^{1/2}\bGamma^{-1}\mathcal{A}(\bGamma^{-1}\cdot\vecF+\grad\cdot\bGamma^{-1}) +\mathrm{noise}$ & Eqs.~\eqref{eq:RT_overdamped_final}, \eqref{eq:RT_overdamped_unit}\\
        \hline
    \end{tblr}
    \caption{Key formulae in this work. U1--5 are expressions pertaining to the underdamped limit, wherein colloidal inertia is kept, and O1--5\,---\,overdamped, where colloid inertia is eliminated. The results we derive in this work are denoted with ``$*$'' and in our companion Letter~\cite{LETTER} by ``$\dagger$''. In that last column, we list relevant references. U1 is the conventionally written underdamped Lagnevin equation. In the case of a free particle (force $\vecF=\mathbf{0}$), it gives rise to an exponential decay of velocity correlations, U2. However, U1 includes only steady-Stokes-flow friction, omitting the Basset memory term (the diffusion of vorticity through the fluid as a particle moves) and the added mass (the momentum of the displaced fluid), which are included in the fluctuating Basset–Boussinesq–Oseen equation (FBBOE), U3. Indeed, it is important to retain the Basset term as velocity autocorrelations decay as a power law with time, U4. In this work, we generalize the BBO to more general geometries using the Lorentz reciprocal theorem (RT), taking the form of U5 where $\cal{A}$ is a geometry-dependent linear memory operator. We next show how Basset memory carries over to the overdamped limit: The conventional overdamped Langevin equation corresponding to U1 is O1. The positional autocorrelation in a trap ($\vecF=-\vecR$, where $\vecR$ is the particle position) decays exponentially, O2. Once the Basset memory term is retained, the corresponding overdamped Basset-Langevin equation (OBLE) to U3 is O3. Indeed, retaining the Basset memory shows that, in fact, positional autocorrelations in the overdamped limit also decay as a power law, O4. We finally derive the corresponding overdamped limit to the RT result U5, given by O5.}
    \label{tab:summary}
\end{table*}

The structure of the remainder of the paper is as follows: 
Via an illustrative example of a spherical colloid in an unbounded fluid, in Sec.~\ref{sec:colloid} we first present how hydrodynamic memory carries over from the FBBOE to the overdamped limit. We recall the special cases of a free and harmonically trapped colloidal particle, showcasing how exponential relaxation crosses over to power-law relaxation even in the overdamped limit. This finding has been largely explored in our companion letter~\cite{LETTER}. Here, this serves as motivating background to the derivation of the two main results of this work: In Sec.~\ref{sec:underdamped}, we derive our first result\,---\,the generalization of the FBBOE to arbitrary geometries and flows\,---\,using the RT, wherein hydrodynamic memory encodes the flow field and the geometry's Green's function. In Sec.~\ref{sec:overdamped} we derive our second result\,---\,the colloid-inertia-less limit corresponding to the general RT result of Sec.~\ref{sec:underdamped}. Finally, in Sec.~\ref{sec:discussion}, we discuss and summarize our results, and propose experimental setups where such an approach would be relevant. Additional technical details are provided as follows: In Appendix~\ref{appendix:BBO}, we show how to derive the FBBOE starting from the general RT result of Sec.~\ref{sec:underdamped}. Appendix~\ref{appendix:sqrt-epsilon} gives the scaling arguments underlying the small-parameter scaling of the Basset memory. Appendix~\ref{appendix:overdamped} provides supporting technical details underlying the overdamped reduction of Sec.~\ref{sec:overdamped}.

\section{Example}\label{sec:colloid}

We first investigate a spherical colloid in an unbounded fluid moving at low Reynolds number with no-slip boundary condition. It will serve as an illustrative special case of the equations of motion we will be deriving in later sections. The ideas will exemplify the mathematical machinery of stochastic expansions as well as the physical motivation for retaining a power-law-decaying kernel during the underdamped-to-overdamped crossover. By ``inertialess'' and ``overdamped'' we strictly refer to the elimination of the particle's inertia (due to the sum of the added-mass term and the particle's mass in its equation of motion); of course the hydrodynamic memory arises from host-fluid inertia, which we must not neglect.

The exemplary system we consider in the present section is a spherical colloid whose radius is $a$ and mass density is $\rho_\p$. We immerse it in an unbounded and uniform fluid whose thermal energy is $\kt$, viscosity is $\eta$, and mass density is $\rho_\f$. Suppose the particle is subjected to a spatial-position- ($\vecr$) and time- ($t$) dependent external force, $\vecf(\vecr,t)$, applied by, \eg optical tweezers. At time $t$, denote the particle's (stochastic) velocity as $\vecv_t$ and position as $\vecr_t$. The conventional underdamped Langevin equation for the particle, \ie that which retains particle inertia, reads~\cite{ClercxPRA1992} 
\begin{equation}
    \frac43\pi a^3\rho_\p\frac{\d\vecv_t}{\d t}=-6\pi \eta a \vecv_t+\vecf(\vecr_t,t)+\sqrt{2\kt(6\pi \eta a)}\,\bxi_t.\label{eq:underdamped}
\end{equation}
Equation~\eqref{eq:underdamped} has been originally proposed by Einstein and Langevin as a force balance~\cite{EinsteinZE1907,LangevinCRASP1908}: On the left-hand side (LHS) is the particle inertia, whereas on the right-hand side (RHS) are the Stokes friction, external (deterministic) force, and thermal (fluctuating) force. As the fluid is in local equilibrium, the white noise must satisfy the fluctuation-dissipation theorem~\cite{PathriaBOOK2011}, requiring
\begin{equation}
    \langle \bxi_t\bxi_{t'}\rangle=\delta(t-t')\vecI,\label{eq:white_noise}
\end{equation}
where $\vecI$ is the $3\times3$ unit matrix. Equation~\eqref{eq:underdamped} can be rigorously brought to the overdamped limit~\cite{SchussBOOK2010,GardinerBOOK2009}, \ie where colloid inertia is eliminated for reasons we will elaborate on shortly and in Sec.~\ref{sec:overdamped}. The overdamped Langevin equation reads
\begin{equation}
    \frac{\d\vecr_t}{\d t}=\frac D\kt\vecf(\vecr_t,t)+\sqrt{2D}\,\bxi_t,\label{eq:overdamped}
\end{equation}
where $D=\kt/(6\pi\eta a)$ is the particle's diffusion constant. The simplified form of Eq.~\eqref{eq:overdamped} has been achieved owing to the exponential decay of velocity autocorrelations in time, implied by Eq.~\eqref{eq:underdamped}.

However, it was long established that velocity autocorrelations do not decay exponentially; rather, they decay as a power law in time, $t^{-3/2}$~\cite{AdlerPRL1967,AdlerPRA1970,ErnstPRL1970,HinchJFM1975,BoonPLA1976,BouillerJPF1978,PaulJPA1981}. While Eq.~\eqref{eq:underdamped} is conceptually appealing, since Brownian motion is never steady, the Stokes friction alone does not fully capture the resistance a no-slip colloid experiences in an incompressible flow. As it traverses unsteadily, the momentum of the displaced fluid must be conserved, thereby incurring an ``added mass'' to the particle's inertia. Furthermore, momentum must be transported throughout the rest of the fluid bulk via diffusion of vorticity, giving rise to the so-called Basset memory. These two effects, along with the steady Stokes friction, constitute the so-called Basset-Boussinesq-Oseen resistance force~\cite{LandauBOOK1987}. Thus, overall, the particle follows the FBBOE~\cite{ClercxPRA1992},
\begin{align}
    \frac43\pi a^3\rho_\p\frac{\d\vecv_t}{\d t}=&-6\pi \eta a \vecv_t-\frac23\pi a^3\rho_\f\frac{\d\vecv_t}{\d t}\nonumber\\& -6a^2\sqrt{\pi\rho_\f\eta}\int_{-\infty}^t\frac{\d \vecv_{t'}}{\sqrt{t-t'}}\nonumber\\&+\vecf(\vecr_t,t)+\sqrt{2\kt(6\pi \eta a)}\,\bta_t,\label{eq:FBBOE}
\end{align}
where, to satisfy fluctuation-dissipation theorem~\cite{HaugeJSP1973}, the noise is colored,
\begin{equation}
    \langle \bta_t\bta_{t'}\rangle=\left[\delta(t-t')-\frac14\sqrt{\frac{\rho_\f a^2}{\pi\eta}}\frac{1}{|t-t'|^{3/2}}\right]\vecI.\label{eq:BBO_noise}
\end{equation}

Here, the full equation of motion for the colloid is known in closed form. The first result of this work is the extension of Eq.~\eqref{eq:FBBOE} to general flows, beyond a spherical particle in an unbounded fluid. We devote Sec.~\ref{sec:underdamped} to deriving this result, Eq.~\eqref{eq:RT-dimensional}. The generality of the result comes at the expense of having to know the corresponding Green's function and flow field around the colloid.
The second result, Eq.~\eqref{eq:RT_overdamped_final}, obtained in Sec.~\ref{sec:overdamped}, answers the following question: Is there a colloid-inertia-free (overdamped) limit of Eq.~\eqref{eq:FBBOE}, given that hydrodynamic memory decays as a power law? 

Indeed, an overdamped (inertialess) limit of Eq.~\eqref{eq:FBBOE} exists, as the dimensionless number controlling the elimination of colloid inertia also sets the magnitude of hydrodynamic memory to be small. In the remainder of this section, we recapitulate the main result of our companion Letter~\cite{LETTER}, using the FBBOE to exemplify the coarse-graining procedure described in Sec.~\ref{sec:overdamped}: To leading-order in small inverse Schmidt number, $\mathrm{Sc}^{-1}=\rho_\f D/\eta$, we proceed to derive
\begin{align}
	\frac{\d\vecr_t}{\d t}=&\frac D\kt\vecf(\vecr_t,t)-\frac D\kt\sqrt{\frac{\rho_\f a^2}{\pi\eta}}\int_{-\infty}^t\frac{\d \vecf(\vecr_{t'},t')}{\sqrt{t-t'}}\nonumber\\&+\sqrt{2D}\,\bxi_t+\sqrt{2D}\sqrt{\frac{\rho_\f a^2}{\pi\eta}}\frac18\int_{-\infty}^\infty\frac{\bxi_{t'}\d t'}{|t-t'|^{3/2}},\label{eq:OBLE}
\end{align}
which we call the overdamped Basset-Langevin equation (OBLE). 

\subsection{Nondimensional equations}

We first nondimensionalize Eq.~\eqref{eq:FBBOE} to identify the small parameter orchestrating the elimination of colloid inertia: The conventional overdamped Langevin equation, Eq.~\eqref{eq:overdamped}, describes a Brownian motion (diffusion constant $D$) under a force $\vecf$ that varies over a lengthscale we denote $w$. We furthermore assume that, if the force is time dependent, it is varies over times comparable to the colloid's diffusive landscape-exploration time, $w^2/D$. Since the particle size does not appear in Eq.~\eqref{eq:overdamped} other than within $D$, $w$ is the relevant lengthscale. (For a free particle, $\vecf=\mathbf0$, one may choose $w=a$ as unbounded diffusion is scale free.) For a time-independent conservative force, by the equipartition theorem~\cite{PathriaBOOK2011}, $\lim_{t\to\infty}\langle\vecr_t\vecf(\vecr_t)\rangle=-\kt\vecI$, which implies that the force magnitude scales as $\kt/w$.  Thus, we define the following dimensionless quantities:
\begin{gather}
    T:=\frac{D}{w^2}t,\quad\vecR:=\frac1 w\vecr,\quad\vecV:=\frac wD\vecv,\nonumber\\\quad\vecF:=\frac w\kt\vecf,\quad(\bTa,\bXi):=\frac{w}{D^{1/2}}(\bta,\bxi).\label{eq:nondim}
\end{gather}
We note that the velocity seems as if it has been set by the diffusion constant $D$, despite Brownian motion not posing a well-defined velocity~\cite{SchussBOOK2010}. In fact, the velocity has been rescaled according to $(D/\kt)\times(\kt/w)$, which is the magnitude of a Stokes-flow response to force\,---\,mobility $D/\kt=1/(6\pi\eta a)$ times force magnitude $\kt/w$.

In terms of these dimensionless variables, Eq.~\eqref{eq:FBBOE} becomes
\begin{equation}
    \kappa\epsilon\frac{\d\vecV_T}{\d T}=-\vecV_T-\epsilon^{1/2}\int_{-\infty}^T\frac{\d \vecV_T'}{\sqrt{T-T'}}+\vecF(\vecR_T,T) +\sqrt{2}\,\bTa_T,\label{eq:noisy_BBO_nondim}
\end{equation}
with the nondimensional (colored) noise
\begin{equation}
    \langle \bTa_T\bTa_{T'}\rangle=\left[\delta(T-T')-\frac{\epsilon^{1/2}}4\frac1{|T-T'|^{3/2}}\right]\vecI.\label{eq:BBO_noise_nondim}
\end{equation}
Equations~\eqref{eq:noisy_BBO_nondim} and~\eqref{eq:BBO_noise_nondim} are the starting point for deriving the overdamped limit, Eq.~\eqref{eq:OBLE}. In Eqs.~\eqref{eq:noisy_BBO_nondim} and~\eqref{eq:BBO_noise_nondim}, we identified two parameters. One is set by the ratio among colloid and fluid densities,
\begin{equation}
    \kappa:=\frac{2\pi}9\left(\frac{\rho_\p}{\rho_\f}+\frac12\right).\label{eq:kappa}
\end{equation}
Often experiments utilize density-matched colloids ($\rho_\p=\rho_\f$) to eliminate gravitational effects, in which case $\kappa=\pi/3$. Thus, we regard $\kappa$ as an order-$1$ parameter. The other, more important parameter is
\begin{equation}
    \epsilon:=\frac1\pi\frac{\rho_\f D}{\eta}\frac{a^2}{w^2},\label{eq:epsilon}
\end{equation}
which is related to the (inverse) Schmidt number, $\mathrm{Sc}^{-1}=\rho_\f D/\eta$, comparing the diffusivity of the colloid with that of momentum within the fluid. For a micron-sized colloid in water at room temperature, $\mathrm{Sc}^{-1}\sim10^{-6}$. Indeed, if momentum diffuses quickly through the fluid compared to how fast the particle moves, the Basset memory would be small in magnitude during the timescale it takes the particle to explore $\vecf(\vecr,t)$. 

Even if the confinement is of a micron-sized colloid is extremely tight (\eg $w\sim10^{-9}\,\mathrm{m}$ versus $a\sim10^{-6}\,\mathrm{m}$ as in the experiments of Ref.~\cite{FranoschNATURE2011}), since $\mathrm{Sc}^{-1}$ is so small, we proceed under the reasonable assumption that $\mathrm{Sc}\gg a/w$. This invites a small-$\epsilon$ expansion in Eq.~\eqref{eq:noisy_BBO_nondim} and~\eqref{eq:BBO_noise_nondim}. Note that the colloid inertia scales as $\epsilon$, whereas Basset memory scales as $\epsilon^{1/2}$, which means that inertia is eliminated prior to the memory. Therefore, since there is a small parameter in front of the particle inertial term, the overdamped limit exists, but Basset memory survives elimination of the colloidal velocity in~\eqref{eq:noisy_BBO_nondim}.

It will be most direct to obtain the overdamped limit by Fourier transforming Eq.~\eqref{eq:noisy_BBO_nondim} in time. This step is natural because (i) other than the force, which may be nonlinear in $\vecR_T$, all other terms are linear in $\vecV_T$ and $\bTa_T$, and (ii) the noise (Eq.~\eqref{eq:BBO_noise_nondim}) diagonalizes in Fourier space. (In Sec.~\ref{sec:overdamped}, we show a more general procedure based on stochastic Taylor expansions~\cite{KloedenBOOK1992,SchussBOOK2010,GardinerBOOK2009}.) We will be working with the convention $\tilde X(\Omega)=\int_{-\infty}^\infty X_Te^{-\i\Omega T}\d T$ for a stochastic time-dependent quantity $X_T$. In the following, we will make use of the properties: $\int_{-\infty}^\infty e^{-\i\Omega T}\d \vecV_T=\i\Omega\tilde \vecV(\Omega)$, $\int_0^\infty e^{-\i\Omega T}T^{-1/2}\d T=(\pi/2)^{1/2}|\Omega|^{-1/2}(1-\i\sgn\Omega )$, and $\int_{-\infty}^{\infty}e^{-\i\Omega T}|T|^{-3/2}\d T=-2(2\pi|\Omega|)^{1/2}$, along with the stochastic-differential notation $(\d \vecV_T/\d T)\d T=\d\vecV_T$. Upon Fourier transforming Eq.~\eqref{eq:noisy_BBO_nondim}, we find
\begin{align}
    \i\kappa\epsilon\Omega\tilde \vecV(\Omega)=&-\tilde\vecV(\Omega)-\sqrt{\frac{\pi\epsilon|\Omega|}2}(1+\i\sgn\Omega )\tilde\vecV (\Omega)\nonumber\\&+\int_{-\infty}^\infty\vecF(\vecR_T,T)e^{-\i\Omega T}\d T+\sqrt{2}\tilde\bTa(\Omega),\label{eq:noisy_BBO_nondim_FT}
\end{align}
with the Fourier-transformed nondimensional (colored) noise
\begin{equation}
    \langle \tilde\bTa(\Omega)\tilde\bTa(\Omega')\rangle=2\pi\delta(\Omega+\Omega')\left(1+\sqrt{\frac{\pi\epsilon|\Omega|}2}\right)\vecI.\label{eq:BBO_noise_nondim_FT}
\end{equation}
The fluctuation-dissipation theorem holds, which can be seen by comparing the real part of the velocity prefactor on the RHS of Eq.~\eqref{eq:noisy_BBO_nondim_FT} and the noise amplitude in Eq.~\eqref{eq:BBO_noise_nondim_FT}. Prior to deriving the OBLE, we first motivate deriving the hydrodynamic corrections to the overdamped limit by two examples\,---\,the free colloid and a colloid within a harmonic trap. The examples will show that hydrodynamic corrections appear in position-related observables as well, thereby implying that memory may appear in the overdamped (velocity-eliminated) limit.

\subsection{Example: Free colloid}\label{sec:free}

Consider the much-studied case of a free particle, $\vecf(\vecr,t)=\boldsymbol{0}$. Equation~\eqref{eq:noisy_BBO_nondim_FT} becomes
\begin{align}
    \i\kappa\epsilon\Omega\tilde \vecV(\Omega)=&-\tilde\vecV(\Omega)-\sqrt{\frac{\pi\epsilon|\Omega|}2}(1+\i\sgn\Omega )\tilde\vecV (\Omega)\nonumber\\&+\sqrt{2}\tilde\bTa(\Omega).
\end{align}
Since the problem of free diffusion is scale free, $w\sim (Dt)^{1/2}$; for concreteness, we consider the timescales it takes for a particle to move a distance of about its own size, $w=a$. Upon isolating $\tilde\vecV(\Omega)$ and using Eq.~\eqref{eq:BBO_noise_nondim_FT}, we obtain
\begin{widetext}
\begin{equation}
    \langle\tilde\vecV(\Omega)\tilde\vecV(\Omega')\rangle=\frac{2(1+\sqrt{\pi\epsilon|\Omega|/2})}{(1+\sqrt{\pi\epsilon|\Omega|/2})^2+\epsilon|\Omega|(\sqrt{\pi/2}+\kappa\sqrt{\epsilon|\Omega|})^2}2\pi\delta(\Omega+\Omega')\vecI.\label{eq:VV_free_FT}
\end{equation}
\end{widetext}

Inverting the Fourier transforms of Eq.~\eqref{eq:VV_free_FT} with $\epsilon\ll1$, we recover the velocity autocorrelation function
\begin{equation}
    \langle\vecV_T\vecV_0\rangle=\left[\delta(T)+\frac{\epsilon^{1/2}}{2T^{3/2}}+\mathcal{O}(\epsilon)\right]\vecI.\label{eq:VV_free}
\end{equation}
This is a well known result~\cite{ErnstPRL1970,HinchJFM1975}, showing how correlations decay as a result of the momentum diffusion into the surrounding fluid much beyond the exponential decay predicted by Eq.~\eqref{eq:underdamped} (with $\vecf=\mathbf0$), $\langle\vecV_T\vecV_0\rangle= e^{-T/(\kappa\epsilon)}\vecI/(\kappa\epsilon)\to\delta(T)\vecI$.
Indeed, Eq.~\eqref{eq:underdamped} is not a hydrodynamically consistent model of underdamped colloidal motion. Namely, if one is after the underdamped model, the fluid's added mass is comparable to the colloid's inertia, and even more so, the Basset memory dominates over colloid inertia at late times. Both contributions to the colloid's effective inertia vanish exponentially quickly whereas hydrodynamic memory persists as a power law.

More pertinent to this work is the mean-squared displacement, obtained by integrating Eq.~\eqref{eq:VV_free},
\begin{equation}
    \langle(\vecR_T-\vecR_0)(\vecR_T-\vecR_0)\rangle=2T\left[1-\frac{2\epsilon^{1/2}}{T^{1/2}}+\calO(\epsilon)\right]\vecI.\label{eq:MSD_free}
\end{equation}
This result has also been obtained in the literature in the past, both theoretically~\cite{WeitzPRL1989} and experimentally~\cite{LukicPRL2005,HuangNP2011}. Equation~\eqref{eq:MSD_free} combines the elements we are after\,---\,the expected leading-order effect from overdamped motion, and a hydrodynamic-memory correction.
In this scale-free case, the hydrodynamic corrections never dominate in Eq.~\eqref{eq:MSD_free}. This is so because the free particle undergoes unbounded diffusion. These corrections decay rather slowly nonetheless, as just $T^{-1/2}$, meaning that the approach to the asymptote is very slow~\cite{WeitzPRL1989}. Yet, if a confining potential is introduced, we posit that Stokesian exponential relaxation due to a force occurs faster than the Basset power-law relaxation. Thus, at truly late times and sufficient sensistivity, relaxation due to hydrodynamic memory in the overdamped limit will still prevail. To see this, we next consider the example of a trapped colloid.

\subsection{Example: Trapped colloid}\label{sec:trap}

We now turn to the case of a trapped particle, $\vecf(\vecr,t)=-\kt\vecr/w^2$. In this case, Eq.~\eqref{eq:noisy_BBO_nondim_FT} becomes
\begin{align}
    -\kappa\epsilon\Omega^2\tilde \vecR(\Omega)=&-\i\Omega\tilde\vecR(\Omega)+\sqrt{\frac{\pi\epsilon|\Omega|^3}2}(1-\i\sgn\Omega)\tilde\vecR (\Omega)\nonumber\\&-\tilde\vecR(\Omega)+\sqrt{2}\tilde\bTa(\Omega),
\end{align}
where we already replaced $\tilde\vecV(\Omega)=\i\Omega\tilde \vecR(\Omega)$. Upon isolating $\tilde\vecR(\Omega)$ and using Eq.~\eqref{eq:BBO_noise_nondim_FT}, we obtain
\begin{widetext}
\begin{equation}
    \langle\tilde\vecR(\Omega)\tilde\vecR(\Omega')\rangle=\frac{2(1+\sqrt{\pi\epsilon|\Omega|/2})}{(1-\sqrt{\pi\epsilon|\Omega|^3/2}-\kappa\epsilon\Omega^2)^2+\Omega^2(1+\sqrt{\pi\epsilon|\Omega|/2})^2}2\pi\delta(\Omega+\Omega')\vecI.
    \label{eq:RR-trap-full}
\end{equation}
\end{widetext}
Inverting the Fourier transforms of Eq.~\eqref{eq:RR-trap-full}, we obtain the position autocorrelation function to leading order in $\epsilon\ll1$ and $T\gg1$,
\begin{equation}
    \langle\vecR_T\vecR_0\rangle=\left[e^{-T}-\frac{\epsilon^{1/2}}{2T^{3/2}}+\calO(\epsilon)+\calO\left(\frac{\epsilon^{1/2}}{T^{5/2}}\right)\right]\vecI.\label{eq:RR_trap}
\end{equation}
In Eq.~\eqref{eq:RR_trap}, the first term is the conventional, exponential relaxation for an overdamped Brownian particle in a harmonic potential, and the second term is a correction due to hydrodynamic memory. We observe that the hydrodynamics gives rise to an anticorrelation\,---\,a particle further away from the trap's minimum would experience greater forces and, as a result, the Basset memory would reduce the friction, thereby increasing the chance for a particle to ``overshoot'' and appear further out on the opposite side of the trap compared to the memoryless case.

\begin{figure*}[t]
    \centering
    \includegraphics[width=0.99\linewidth]{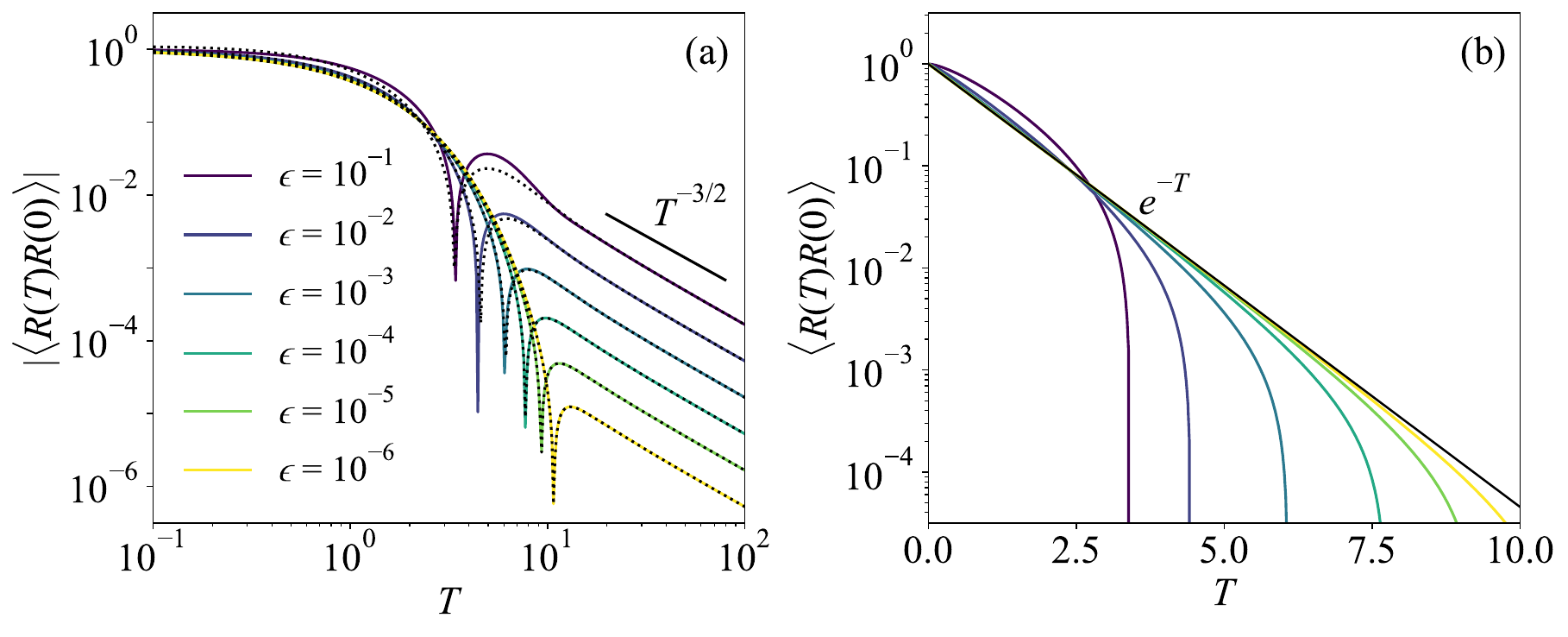}
    \caption{Absolute value of the positional autocorrelation \(|\langle R_T R_0\rangle|\) versus time for a harmonically trapped spherical colloid on  (a) log-log and (b) log-linear plots. Solid curves show the full spectral result, Eq.~\eqref{eq:RR-trap-full}; dashed curves show the leading-order \(\sqrt\epsilon\) approximation, Eq.~\eqref{eq:RR_trap}. The $T^{-3/2}$ guide in (a) highlights the late-time hydrodynamic-memory tail. The $e^{-T}$ guide in (b) shows that for smaller $\epsilon$ the exponential fits remain accurate for a longer time.}
    \label{fig:trap-linResp}
\end{figure*}

Equation~\eqref{eq:RR_trap} demonstrates the role of hydrodynamic memory in overdamped relaxational dynamics. The conventional overdamped Langevin equation (Eq.~\eqref{eq:overdamped}) predicts simply $\langle\vecR_T\vecR_0\rangle=e^{-T}\vecI$. Indeed, this is the leading-order behavior in Eq.~\eqref{eq:RR_trap}. However, when sufficient experimental resolution is available, the relaxation crosses-over from exponential steady-Stokes-friction dominated dynamics to power-law Basset-memory dominated dynamics~\cite{FranoschNATURE2011}. (A more detailed comparison against experiments can be found in our companion work~\cite{LETTER}.) Figure~\ref{fig:trap-linResp} illustrates this crossover by comparing the full linear-response result with the leading-order $\epsilon^{1/2}$ expression. 

This suggests that even in the overdamped limit, hydrodynamic correlations always dominate at sufficiently late time scales. The reason why such effects have not been widely reported in the past is because, for sufficiently small $\epsilon$, Fig.~\ref{fig:trap-linResp}(b) suggests that the exponential decay is a good approximation for a few decades in time, so the power law $3/2$ of Fig.~\ref{fig:trap-linResp}(a) is no longer detectable or statistically significant experimentally. Our work here and in our companion paper~\cite{LETTER} suggest that the power-law correlations are nonetheless always present and dominate at late times; it is simply a matter of having a sensitive enough measurement. This example serves as a motivation to derive the OBLE that includes these hydrodynamic-memory corrections.

\subsection{Overdamped limit}

Motivated to obtain the general overdamped limit, we return to Eq.~\eqref{eq:FBBOE} with the general force and derive the corresponding OBLE. Equation~\eqref{eq:noisy_BBO_nondim_FT} is an algebraic equation in $\tilde\vecV(\Omega)$, which we may thus isolate. To order $\epsilon^{1/2}$, we eliminate the inertia terms and express $\tilde\vecV(\Omega)$ as
\begin{align}
    \tilde \vecV(\Omega)=&\left[1-\sqrt{\frac{\pi\epsilon|\Omega|}2}(1+\i\sgn\Omega)\right]\nonumber\\&\times\int_{-\infty}^\infty\vecF(\vecR_T,T)e^{-\i\Omega T}\d T+\sqrt2\tilde\bTa'(\Omega)+\mathcal{O}(\epsilon),\label{eq:V_nondim_FT_interrim}
\end{align}
where we defined $\tilde\bTa'(\Omega)=[1-(\pi\epsilon|\Omega|/2)^{1/2}(1+\i\sgn\Omega)]\tilde\bTa(\Omega)$. To order $\epsilon^{1/2}$, this noise satisfies
\begin{equation}
    \langle \tilde\bTa'(\Omega)\tilde\bTa'(\Omega')\rangle=2\pi\delta(\Omega+\Omega')\left[1-\sqrt{\frac{\pi\epsilon|\Omega|}2}+\mathcal{O}(\epsilon)\right]\vecI.\label{eq:V_noise_nondim_FT_interrim}
\end{equation}
Recalling that the Fourier transform of a (nondimenensionalized) white noise is
\begin{equation}
    \langle \tilde\bXi(\Omega)\tilde\bXi(\Omega')\rangle=2\pi\delta(\Omega+\Omega')\vecI,\label{eq:white_noise_nondim_FT}
\end{equation}
we may replace $\tilde\bTa'(\Omega)$ in Eq.~\eqref{eq:V_nondim_FT_interrim} with white noise as follows:
\begin{widetext}
\begin{equation}
    \tilde \vecV(\Omega)=\left[1-\sqrt{\frac{\pi\epsilon|\Omega|}2}(1+\i\sgn\Omega)\right]\int_{-\infty}^\infty\vecF(\vecR_T,T)e^{-\i\Omega T}\d T+\sqrt2\left(1-\frac12\sqrt{\frac{\pi\epsilon|\Omega|}2}\right)\tilde\bXi(\Omega).\label{eq:V_nondim_FT}
\end{equation}
One may confirm that Eqs.~\eqref{eq:VV_free}, \eqref{eq:MSD_free}, and~\eqref{eq:RR_trap} are recovered directly from Eq.~\eqref{eq:V_nondim_FT}.

An inverse Fourier transform yields
\begin{equation}
    \vecV_T=\vecF(\vecR_T,T)-\epsilon^{1/2}\int_{-\infty}^T\frac{\d \vecF(\vecR_T',T')}{\sqrt{T-T'}}+\sqrt2\bXi_T+\frac{\sqrt{2\epsilon}}8\int_{-\infty}^\infty\frac{\bXi_{T'}\d T'}{|T-T'|^{3/2}},\label{eq:V_nondim_interrim}
\end{equation}
where the (dimensionless) white noise satisfies
\begin{equation}
    \langle\bXi_T\bXi_{T'}\rangle=\delta(T-T')\vecI.\label{eq:nondim_white_noise}
\end{equation}
Equation~\eqref{eq:V_nondim_interrim} is a stochastic integrodifferential equation for $\vecR_T$; upon reintroducing units, we recover Eq.~\eqref{eq:OBLE}.

This is not yet a closed-form Langevin equation, as it involves the increment of change in force along a trajectory, $\d\vecF (\vecR_T,T)$. This is a consequence of the Basset memory, introducing a dependence on past accelerations. We simplify this term using the It\^o lemma~\cite{SchussBOOK2010}, where $\d \vecR_T$ satisfies Eq.~\eqref{eq:V_nondim_interrim},
\begin{equation}
    \frac{\d \vecF(\vecR_{T},T)}{\d T}=\frac{\partial \vecF(\vecR_{T},T)}{\partial T}+[\vecF(\vecR_T,T)\cdot\grad ]\vecF(\vecR_T,T)+\sqrt2(\bXi_T\cdot\grad )\vecF(\vecR_T,T)+D\nabla^2\vecF(\vecR_T,T)+\mathcal{O}(\epsilon^{1/2}).
\end{equation}
With that, we find the overdamped limit of Eq.~\eqref{eq:noisy_BBO_nondim},
\begin{eqnarray}
	\frac{\d\vecR_T}{\d T}&=&\vecF(\vecR_T,T)-\sqrt\epsilon\int_{-\infty}^T\frac{[\vecF(\vecR_{T'},T')\cdot\grad]\vecF(\vecR_{T'},T')+\nabla^2\vecF(\vecR_{T'},T')+\partial\vecF(\vecR_{T'},T')/\partial T'}{\sqrt{T-T'}}\d T'
	\nonumber\\&&+\sqrt2\bXi_T+\sqrt{2\epsilon}\left[\frac18\int_{-\infty}^\infty\frac{\bXi_{T'}\d T'}{|T-T'|^{3/2}}-\int_{-\infty}^T\frac{(\bXi_{T'}\cdot\grad)\vecF(\vecR_{T'},T')}{\sqrt{T-T'}}\d T'\right].\label{eq:V_nondim}
\end{eqnarray}
\end{widetext}
Out of the three contributions to friction in the BBO equation\,---\,(steady) Stokes drag, Basset memory, and added mass\,---\,Stokes drag is the dominant contribution to the colloidal dynamics in Eq.~\eqref{eq:V_nondim}. Upon eliminating the colloid's inertia, since the (displaced) fluid has a similar density ($\kappa\sim1$), the added-mass term has been eliminated alongside it. Ultimately, an order-$\epsilon^{1/2}$ correction arose from the hydrodynamic memory.

In this section, we considered a spherical colloid in an unbounded fluid as an illustrative example of (i)  the importance of hydrodynamic memory in determining late-time asymptotics and (ii) the existence of a well-defined overdamped (colloid-inertia-less) limit of an equation with power-law memory. Our aim in this work is to include hydrodynamic memory in a very general setting.

\section{Generalized underdamped dynamics}\label{sec:underdamped}

In this section, we derive the underdamped equation of motion of a no-slip particle in a fluctuating viscous fluid in a general confinement, Eq.~\eqref{eq:RT-Langevin} below. Specializing to the spherical unbounded case, the equation reduces to the BBO Eq.~\eqref{eq:BBO_noise_nondim} used in Sec.~\ref{sec:colloid}; see Appendix~\ref{appendix:BBO}. The derivation follows the  reciprocal theorem~\cite{TurkJFM2025} applied to the Navier-Stokes equations, including the particle inertia, while retaining the thermal (Brownian) and fluid-inertial effects of a no-slip colloidal particle.

Consider an incompressible flow of a Newtonian fluid occupying a volume $V$, with boundary made of the particle surface $(S_\p)$, possible fixed boundaries $(S_\b)$, and a surface at infinity $(S_\infty)$, \ie $\partial V=S_\p\cup S_\b\cup S_\infty$. The unit normal $\vecn$ points from the particle or boundary into the fluid volume. The fluid has velocity $\vecu(\vecr,t)$, pressure $p(\vecr,t)$, and fluctuating bulk thermal force density $\bbeta(\vecr,t)$. In the limit of low particle Reynolds number, we retain the unsteady Stokes term but neglect the convective inertia~\cite{HaugeJSP1973},
\begin{align}
    \grad\cdot\vecu(\vecr,t)=&0,\nonumber\\
    \rho_\f \frac{\partial\vecu(\vecr,t)}{\partial t}=&\grad\cdot\bsigma(\vecr,t)+\bbeta(\vecr,t),
    \label{eq:fluctuating-stokes}
\end{align}
where the fluid stress tensor $\bsigma(\vecr,t)$ is
\begin{equation}
    \bsigma(\vecr,t)=-p(\vecr,t)\vecI+\eta\left\{\grad\vecu(\vecr,t)+[\grad\vecu(\vecr,t)]^\T\right\},
    \label{eq:fluid-stress}
\end{equation}
with $\rho_\f$ the fluid's density and $\eta$ the fluid's viscosity. The fluid is locally at thermal equilibrium, so the fluctuating hydrodynamic force $\bbeta(\vecr,t)$ must be such that the fluctuation-dissipation theorem holds with a temperature $\kt$. Namely, its contraction with a test flow satisfies~\cite{HaugeJSP1973}
\begin{widetext}
\begin{equation}
    \left\langle
    \int_V \vecy(\vecr,t)\cdot\bbeta(\vecr,t)\d\vecr
    \int_V \bbeta(\vecr',t')\cdot\vecy(\vecr',t')\d V'
    \right\rangle
    =2\kt \left\{\frac{\eta}{2}\int_V\left\Vert\grad\vecy(\vecr,t)+[\grad\vecy(\vecr,t)]^\T\right\Vert^2\d V\right\}\delta(t-t'),
    \label{eq:fluid-FDT}
\end{equation}
\end{widetext}
where $\vecy(\vecr,t)$ is an incompressible test field satisfying the same boundary conditions on $\partial V$. The magnitude of fluctuations (the argument of the curly braces on the RHS in Eq.~\eqref{eq:fluid-FDT}) is the positive steady-Stokes dissipation associated with the test flow, and $\Vert \vecO\Vert^2=\vecO:\vecO$ is the Frobenius matrix norm of a matrix $\vecO$. 

To make use of the RT, we compare the actual fluctuating flow $\vecu(\vecr,t)$ to an auxiliary zero-temperature steady Stokes flow under a test force, denoted by hats, at the same (frozen) instantaneous particle configuration, $\vecr_t$. The reciprocal theorem for the unsteady Stokes equation gives~\cite{TurkJFM2025}
\begin{align}
    \int_{S_\p}\vecn(\vecr)\cdot\hat{\bsigma}(\vecr)\cdot\vecu(\vecr,t)\d S=&
    \int_{S_\p}\vecn(\vecr)\cdot\boldsymbol{\sigma}(\vecr,t)\cdot\hat{\vecu}(\vecr)\d S
    \nonumber\\&-\int_V \bbeta(\vecr,t)\cdot\hat{\vecu}(\vecr)\d V
    \nonumber\\&+\rho_\f\int_V \frac{\partial\vecu(\vecr,t)}{\partial t}\cdot\hat{\vecu}(\vecr)\d V .
    \label{eq:RT}
\end{align}
The actual and auxiliary fields satisfy the same no-penetration, no-slip, or stress-free conditions on the fixed-boundary surface $S_\b$, and the auxiliary Stokes flow is assumed to decay sufficiently rapidly on the far-field surface $S_\infty$, so these surface terms vanish under the standard reciprocal-theorem assumptions. 

We denote the particle's time-dependent position $\vecr_t$ and solid angle $\bomega_t$, and the corresponding linear velocity $\vecv_t$ and angular velocity $\bnu_t$. On the particle surface, the boundary condition is
\begin{equation}
    \vecu(\vecr,t)=\vecv_t+\bnu_t\times(\vecr-\vecr_t),
    \qquad \vecr\in S_\p,
    \label{eq:rigid-boundary}
\end{equation}
with no interfacial slip. Newton's equations for the force and torque are
\begin{align}
    m\frac{\d \vecv_t}{\d t}=&\vecf(\vecr_t,\bomega_t,t)+\int_{S_\p}\vecn(\vecr)\cdot\bsigma(\vecr,t)\d S,\label{eq:newton-force}\\
    \vecj\cdot\frac{\d\bnu_t}{\d t}=&\btau(\vecr_t,\bomega_t,t)+\int_{S_\p}(\vecr-\vecr_t)\times[\vecn(\vecr)\cdot\bsigma(\vecr,t)]\d S,
    \label{eq:newton-torque}
\end{align}
where $\vecf$ and $\btau$ are external force and torque, and $m$ and $\vecj$ are the particle mass and moment-of-inertia tensor.
For bookkeeping purposes of both translational and rotational components, let $\und\vecr_t=(\vecr_t,a\bomega_t)^\T$, $\und\vecv_t=(\vecv_t,a\bnu_t)^\T$, $\und\vecf=(\vecf,\btau/a)^\T$, and $\und{\und\vecm}=\operatorname{diag}(m\vecI,\vecj/a^2)$. 

Note that boldface vectors and tensors are of size $3\times1$ and $3\times3$, respectively; the added underscores indicates that we grouped positional and rotational vectors and tensors, so they are of size $6\times1$ and $6\times6$ overall.

The auxiliary steady-Stokes problem defines the grand friction tensor, connecting the steady auxiliary particle velocity $\hat\vecv$ with auxiliary external force $\hat\vecf$,
\begin{equation}
    \und{\hat\vecf}=\und{\und\bgamma}(\und\vecr_t)\cdot\und{\hat\vecv},
    \label{eq:grand-friction}
\end{equation}
where $\und{\und\bgamma}$ is the ordinary steady Stokes grand friction tensor, at the particles instantaneous configuration $\vecr_t$; it is thus solely dependent on geometry. We substitute  Eqs.~\eqref{eq:rigid-boundary}--\eqref{eq:grand-friction} in the particle-surface terms (LHS and first term on the RHS of Eq.~\eqref{eq:RT}) to find
\begin{widetext}
\begin{equation}
    -\left[\und{\und\bgamma}(\und\vecr_t)\cdot\und{\hat\vecv}\right]\cdot\und\vecv_t=
    \left[\und{\und\vecm}\cdot \frac{\d \und\vecv_t}{\d t}-\und\vecf(\und\vecr_t,t)\right]\cdot\und{\hat\vecv}
    -\int \bbeta(\vecr,t)\cdot\hat{\vecu}(\vecr)\d\vecr
    +\rho_\f\int \frac{\partial\vecu(\vecr,t)}{\partial t}\cdot\hat{\vecu}(\vecr)\d\vecr,
    \label{eq:RT_interrim}
\end{equation}
where, since we will not have surface integrals later in this work, we will denote volume integrals by $\d\vecr$ instead of $\d V$ for convenience later.

To simplify the second (noise) term on the RHS of Eq.~\eqref{eq:RT_interrim}, note that for an incompressible flow, $\eta\left\Vert\grad\hat\vecu(\vecr)+[\grad\hat\vecu(\vecr)]^\T\right\Vert^2=2\hat\bsigma(\vecr):[\grad\hat\vecu(\vecr)]$ and furthermore for a steady Stokes flow ($\grad\cdot\hat\bsigma(\vecr)=\mathbf0$), $\hat\bsigma(\vecr):[\grad\hat\vecu(\vecr)]=\grad\cdot[\hat\bsigma(\vecr)\cdot\hat\vecu(\vecr)]$. This allows to bring the RHS of Eq.~\eqref{eq:fluid-FDT} for the auxiliary flow ($\vecy\to\hat\vecu$) into a particle-surface integral, yielding
\begin{equation}
    \left\langle
    \int \hat\vecu(\vecr)\cdot\bbeta(\vecr,t)\d \vecr
    \int \bbeta(\vecr',t')\cdot\hat\vecu(\vecr')\d \vecr'
    \right\rangle
    =2\kt[\und{\hat\vecv}\cdot\und{\und\bgamma}(\und\vecr_t)\cdot\und{\hat\vecv}]\delta(t-t').\label{eq:dissipation-incomp}
\end{equation}
\end{widetext}
To simplify the third term of Eq.~\eqref{eq:RT_interrim}, we employ the linearity of the auxiliary problem to define a $6\times 3$ matrix $\mathsf{k}$ through a Fax\'en-like relation,
\begin{equation}
    \hat{\vecu}(\vecr)
    =
    \mathsf{k}^\T(\und\vecr_t,\vecr)\cdot\und{\hat\vecf}.
    \label{eq:kernel-definition-general}
\end{equation}
Since $\mathsf{k}$ relates to the auxiliary steady Stokes problem, it is solely a function of geometry. For a sphere of radius $a$, this generalized kernel has the translational and rotational blocks $\mathsf{k}(\und\vecr,\vecr')=([1+(a^2/6)\nabla'^2]\vecg(\vecr,\vecr'),(1/2)\grad'\times\vecg(\vecr,\vecr'))^\T$, with $\vecg$ the steady Stokes Green's tensor for the geometry and we used the short-hand $\grad' = \grad_{\vecr'}$.

Substituting Eqs.~\eqref{eq:dissipation-incomp} and~\eqref{eq:kernel-definition-general} into Eq.~\eqref{eq:RT_interrim}, and using the arbitrariness of $\und{\hat\vecv}$, yields the full RT Langevin equation
\begin{align}
    \und{\und\vecm}\cdot\frac{\d\und\vecv_t}{\d t}=&-\und{\und\bgamma}(\und\vecr_t)\cdot\und\vecv_t
    -\rho_\f\und{\und\bgamma}(\und\vecr_t)\cdot
    \int\mathsf{k}(\und\vecr_t,\vecr)\cdot\frac{\partial\vecu(\vecr,t)}{\partial t}\d\vecr\nonumber\\
    &+\und\vecf(\und\vecr_t,t)
    +\sqrt{2\kt\und{\und\bgamma}(\und\vecr_t)}\cdot\und\bxi_t,
    \label{eq:RT-dimensional_almost}
\end{align}
where $\und{\und\bgamma}^{1/2}\cdot\und\bxi_t$ is to be interpreted such that  $\langle[\und{\und\bgamma}^{1/2}(\und\vecr_t)\cdot\und\bxi_t][\und{\und\bgamma}^{1/2}(\und\vecr_{t'})\cdot\und\bxi_{t'}]\rangle=\und{\und\bgamma}(\und\vecr_t)\delta(t-t')$ (\ie since $\und{\und\bgamma}$ is symmetric and positive definite, a square-matrix square root of it exists). The last term is a white noise arising from the steady auxiliary Stokes contraction; after the unsteady fluid response in the volume integral is evaluated, the same equilibrium bath produces the colored particle-level noise required by the full hydrodynamic fluctuation-dissipation theorem~\cite{HaugeJSP1973}. When rotations are suppressed, $\und\vecv_t$ and $\und\vecf$ reduce to the translational $\vecv_t$ and $\vecf$ of Sec.~\ref{sec:colloid}. For a sphere in bulk, in Appendix~\ref{appendix:BBO} we verify that Eq.~\eqref{eq:RT-dimensional_almost} reproduces the FBBOE. Otherwise, Eq.~\eqref{eq:RT-dimensional_almost} is the most general form of a Langevin equation for a particle immersed in a viscous fluid. 

Indeed, the generality of Eq.~\eqref{eq:RT-dimensional_almost} comes at the cost that one must know $\und{\und\bgamma}(\vecr)$ and $\mathsf{k}(\vecr,\vecr')$ of the steady-Stokes problem corresponding to the given geometry (\ie particle and confinement shape), and furthermore the full flow field $\vecu(\vecr,t)$. Supposing these quantities are given via some method~\cite{KimBOOK2013,StakgoldBOOK1979}, we can simplify our result further: Note that the Stokes equation, Eq.~\eqref{eq:fluctuating-stokes}, is (i) strictly linear in the fluid and particle velocities and (ii) the noise satisfies the fluctuation-dissipation theorem (Eqs.~\eqref{eq:fluid-FDT}). Consider first the noiseless limit: Regardless of the specific geometry considered and the corresponding solutions $\mathsf{k}(\vecr;\vecr')$ and $\vecu(\vecr,t)$, in most generality, we are allowed to rewrite the convolution in Eq.~\eqref{eq:RT-dimensional_almost} in the determinstic case as
\begin{equation}
    \int \mathsf{k}(\und\vecr_t,\vecr)\cdot\frac{\partial\vecu(\vecr,t)}{\partial t}\d \vecr:=\int_{-\infty}^t\und{\und\vecb}(\und\vecr_t,t;t')\cdot\und\vecv_{t'}\d t',\label{eq:operator_a_def}
\end{equation}
where $\und{\und\vecb}(\und\vecr,t;t')$ is the ``tensor of weights'' in the linear combination of the velocity history, unique to the considered geometry and flow problem. Note that $\und{\und\vecb}$ can in fact be a linear operator acting on $\vecv_s$, parametrized by the instantaneous $\vecr_t$, $t$, and $s$. Thus, we rewrite Eq.~\eqref{eq:RT-dimensional_almost} as
\begin{align}
    \und{\und\vecm}\cdot\frac{\d\und\vecv_t}{\d t}=&-\und{\und\bgamma}(\und\vecr_t)\cdot\und\vecv_t
    -\rho_\f\und{\und\bgamma}(\und\vecr_t)\cdot\int_{-\infty}^t\und{\und\vecb}(\und\vecr_t,t;t')\cdot\und\vecv_{t'}\d t'\nonumber\\
    &+\und\vecf(\und\vecr_t,t)
    +\sqrt{2\kt\und{\und\bgamma}(\und\vecr_t)}\cdot\und\bta_t,
    \label{eq:RT-dimensional}
\end{align}
where the deterministic part of $\vecu(\vecr,t)$ have been written out explicitly using Eq.~\eqref{eq:operator_a_def}, but its fluctuating part has been absorbed into the last (noise) term. Although we have not prescribed explicitly the fluctuating contribution from the convolution, since fluctuation-dissipation theorem is satisfied~\cite{PathriaBOOK2011,HaugeJSP1973}, we are guaranteed that its autocorrelation must be given by
\begin{widetext}
\begin{equation}
    \langle\und\bta_t\und\bta_{t'}\rangle=\delta(t-t')\und{\und\vecI}+\frac{\rho_\f}2\left\{\sqrt{\und{\und\bgamma}(\und\vecr_t)}\cdot\und{\und\vecb}(\und\vecr_t,t;t')\cdot\sqrt{\und{\und\bgamma}^{-1}(\und\vecr_{t'})}\Theta(t-t')+\left[\sqrt{\und{\und\bgamma}(\und\vecr_{t'})}\cdot\und{\und\vecb}(\und\vecr_{t'},t';t)\cdot\sqrt{\und{\und\bgamma}^{-1}(\und\vecr_t)}\right]^\T\Theta(t'-t)\right\},\label{eq:RT-dimensional_noise}
\end{equation}
\end{widetext}
where $\Theta(x)$ is the Heaviside step function ($\Theta(x>0)=1$ and $\Theta(x<0)=0$).

Equation~\eqref{eq:RT-dimensional} is the final form of our first main result in this work. Although the kernel $\vecb$ requires involved information about the geometry and fluid flow properties (\eg see Appendix~\ref{appendix:BBO}), Eq.~\eqref{eq:RT-dimensional} is nonetheless a closed-form equation of motion for $\vecv_t$. For a low-Reynolds-number flow, this memory kernel encodes all hydrodynamic effects arising from fluid inertia; all other terms are those which appear in the conventional underdamped Langevin equation, Eq.~\eqref{eq:underdamped}. Indeed, upon specifying to spherical colloid in an unbounded fluid, $\bgamma=6\pi\eta a\,\vecI$ and $\vecb(t;s)=\vecI [a(\pi\rho_\f\eta)^{-1/2}(t-s)^{-1/2}(\d/\d s)+ a^2/(9\eta)\delta(t-s)(\d/\d s)]$, Eq.~\eqref{eq:RT-dimensional} reduces to Eq.~\eqref{eq:FBBOE}; see Appendix~\ref{appendix:BBO}.

\section{Generalized overdamped dynamics}\label{sec:overdamped}

\subsection{Nondimensional equations}

Equation~\eqref{eq:RT-dimensional} retains the inertia of the colloid. The analysis in Sec.~\ref{sec:colloid} suggests that particle inertia  may become negligible by the time it takes the colloid to explore the force landscape. To see if this conclusion holds in the most general RT result, we repeat the nondimensionalization of Sec.~\ref{sec:colloid}. For a colloid with a linear dimension $a$, the Stokes friction tensor must scales as $\eta a$, diffusivity as $\kt/(\eta a)$, and mass as $\rho_\p a^3$. At the same time, the external force varies across a distance $w$, which is thus the relevant lengthscale, so the relevant timescale is $w^2/[\kt/(\eta a)]$. Like before, the force magnitude is $\kt/w$, so the velocity is of order $(\kt/w)/(\eta a)$. Next, we nondimensionalize all quantities in Eq.~\eqref{eq:RT-dimensional} as follows:
\begin{gather}
   T:=\frac{\kt}{\eta aw^2}t,\quad\und\vecR:=\frac1w\und\vecr,\quad\und\vecV:=\frac{\eta aw}\kt\und\vecv,\nonumber\\ \und{\und\vecM}:=\frac{1}{\rho_\f a^3}\und{\und\vecm},\quad\und\vecF:=\frac{w}{\kt}\und\vecf,\nonumber\\
   \und{\und\bGamma}:=\frac1{\eta a}\und{\und\bgamma},\quad(\und\bTa,\und\bXi):=\left(\frac{\eta aw^2}{\kt}\right)^{1/2}(\und\bta,\und\bxi).\quad\label{eq:nondimfull}
\end{gather}
For the mass, we assumed that $\rho_\f \sim\rho_\p$ (\cf $\kappa\sim1$ of Eq.~\eqref{eq:kappa}). 

Nondimensionalizing the hydrodynamic kernel would require a problem-specific treatment. As an example, consider an unbounded fluid but an arbitrarily shaped particle. In Appendix~\ref{appendix:sqrt-epsilon} we detail a scaling argument, which holds for both translational and rotational diffusion of the particle: The diffusion of vorticity, $\eta/\rho_\f$, gives rise to the Basset memory length $[(\eta/\rho_\f)\times \eta aw^2/(\kt)]^{1/2}$ during the diffusion time $\eta aw^2/(\kt)$. Thus, with
\begin{gather}
    \mathsf{k}:=\frac{1}{\eta a}\mathsf K,\quad\vecu:=\frac{\eta aw}\kt\vecV,\quad t=\frac{\eta aw^2}\kt T,\nonumber\\\d \vecr:= a\times a\times\sqrt{\frac{\eta^2 aw^2}{\rho_\f\kt}}\d\vecR,
\end{gather}
where the latter is the basset shell volume, we nondimensionalize the kernel as
\begin{align}
    \int \mathsf{k}(\und\vecr_t,\vecr)\cdot\frac{\partial\vecu(\vecr,t)}{\partial t}\d\vecr=&\frac{1}{\rho_\f \eta a}\frac{\kt}{w}\theta\nonumber\\&\times
    \int \mathsf{K}(\und\vecR_T,\vecR)\cdot\frac{\partial\vecU(\vecR,T)}{\partial T}\d \vecR,
\end{align}
where we defined $\theta= (\rho_\f a\kt)^{1/2}/(\eta w)$ for the unbounded-fluid case. In a different geometry (\eg a particle in a channel), one would obtain a different parameter $\theta$. Using
\begin{equation}
    \int \mathsf{K}(\und\vecR_T,\vecR)\cdot\frac{\partial\vecU(\vecR,T)}{\partial T}\d \vecR=\int_{-\infty}^T\und{\und\vecB}(\und\vecR_T,T;T')\cdot\und\vecV_{T'}\d T'
\end{equation}
and Eq.~\eqref{eq:operator_a_def}, we identify the nondimensional kernel as 
\begin{equation}
    \und{\und\vecb}:=\frac{\kt}{\rho_\f \eta aw^2}\theta\und{\und\vecB}.
\end{equation}

With the above dimensionless parameters, Eq.~\eqref{eq:RT-dimensional} becomes
\begin{widetext}
\begin{equation}
    \epsilon\und{\und\vecM}\cdot\frac{\d\und\vecV_T}{\d T}
    :=
    -\und{\und\bGamma}(\und\vecR_T)\cdot\und\vecV_T
    -\theta\,\und{\und\bGamma}(\und\vecR_T)\cdot
    \int_{-\infty}^T\und{\und\vecB}(\und\vecR_T,T;T')\cdot\und\vecV_{T'}\d T'
    +\und\vecF(\und\vecR_T,T)
    +\sqrt{2\und{\und\bGamma}(\und\vecR_T)}\cdot\und\bTa_T,
    \label{eq:RT-Langevin}
\end{equation}
where the noise is
\begin{align}
    \langle\und\bTa_T\und\bTa_{T'}\rangle=\delta(T-T')\und{\und\vecI}+\frac\theta2&\left\{\sqrt{\und{\und\bGamma}(\und\vecR_T)}\cdot\und{\und\vecB}(\und\vecR_T,T;T')\cdot\sqrt{\und{\und\bGamma}^{-1}(\und\vecR_{T'})}\Theta(T-T')\right.\nonumber\\&+\left.\left[\sqrt{\und{\und\bGamma}(\und\vecR_{T'})}\cdot\und{\und\vecB}(\und\vecR_{T'},T';T)\cdot\sqrt{\und{\und\bGamma}^{-1}(\und\vecR_T)}\right]^\T\Theta(T'-T)\right\}.\label{eq:RT-BBO_noise}
\end{align}
\end{widetext}
Once again, the parameter 
\begin{equation}
    \epsilon=\frac{\rho_\f[\kt/(\eta a)]}{\eta}\frac{a^2}{w^2}
\end{equation} 
is related to the Schmidt number $\mathrm{Sc}^{-1}=\rho_\f[\kt/(\eta a)]/\eta$ and the ratio of length scales $a/w$; \cf Eq.~\eqref{eq:epsilon}. (Note that $\theta=\epsilon^{1/2}$ in an unbounded fluid.) Here, too, $\epsilon$ controls both the elimination of the inertia and the magnitude of hydrodynamic memory. While inertia scales as $\epsilon$, the $\epsilon^{1/2}$ prefactor follows from the Basset length in the unsteady Stokes response of a particle in an unbounded fluid. As before, the inertial influence of the colloidal particle appears to be smaller in magnitude than the influence of hydrodynamic memory. Thus, the general overdamped reduction, \ie elimination of inertia ($\epsilon\ll1$ expansion), of Eq.~\eqref{eq:RT-Langevin} will yield an overdamped Langevin equation with a correction from the RT-hydrodynamic memory kernel. 

Since a colloidal particle within a fluid in different confinements will exhibit different hydrodynamic memory and therefore scaling with system parameters, in the following we treat $\theta$ in Eqs.~\eqref{eq:RT-Langevin} and~\eqref{eq:RT-BBO_noise} as a parameter which may not scale as $\epsilon^{1/2}$ of the unbounded geometry, but nonetheless satisfies $\epsilon\ll\theta\ll1$; $\theta$ thus represents the small parameter controlling the geometry- and flow-dependent memory. (In the case of an unbounded fluid, we obtain the above $\theta=\epsilon^{1/2}$.) In the next section, we perform the stochastic Taylor-expansions of the general overdamped equation, Eq.~\eqref{eq:RT-Langevin}, wherein we eliminate $\epsilon$ and keep leading-order corrections in $\theta$.
Since all vectors and tensors in Eqs.~\eqref{eq:RT-Langevin} and~\eqref{eq:RT-BBO_noise} are of size $6$ and $6\times6$, respectively, we will drop the underbars for brevity.

\subsection{Overdamped limit}

Since Eq.~\eqref{eq:RT-Langevin} is no longer linear in $\vecR_T$ in any of the terms, eliminating inertia via a Fourier transform as we did in Sec.~\ref{sec:colloid} is not possible. Instead, we will perform small $\epsilon\ll\theta\ll1$ expansions directly. First, we express the colored $\bTa_T$ in terms of the white noise $\bXi_T$, satisfying Eq.~\eqref{eq:nondim_white_noise}, by expanding up to order $\theta$:
\begin{widetext}
\begin{equation}
    \bTa_T=\bXi_T+\frac{\theta}2\sqrt{\bGamma(\vecR_T)}\cdot\int_{-\infty}^T\vecB(\vecR_T,T;S)\cdot\sqrt{\bGamma^{-1}(\vecR_S)}\cdot\bXi_S\d S+\calO(\theta^2).
\end{equation}
One can confirm that its autocorrelation coincides with Eq.~\eqref{eq:RT-BBO_noise} up to order $\theta$. Next, define
\begin{equation}
    \vecQ_{T\leftarrow T'}:=\exp_\T\left[-(\epsilon\vecM)^{-1}\cdot\int_{T'}^T\d T''\bGamma (\vecR_{T''})\right]\cdot (\epsilon\vecM)^{-1},\label{eq:Q-def}
\end{equation}
where $\exp_\T$ is the time-ordered matrix exponential, with which we implicitly solve Eq.~\eqref{eq:RT-Langevin} as
\begin{eqnarray}
    \vecV_T
    &=&\int_{-\infty}^T\d T'\vecQ_{T\leftarrow T'}\cdot\vecF(\vecR_{T'},T')+\int_{-\infty}^{T}\d T'\vecQ_{T\leftarrow T'}\cdot\sqrt{2\bGamma(\vecR_{T'})}\cdot\bXi_{T'}\nonumber\\&&-\theta\int_{-\infty}^T\d T'\vecQ_{T\leftarrow T'}\cdot\bGamma(\vecR_{T'})\cdot\int_{-\infty}^{T'}\d S\vecB(\vecR_{T'},T';S)\cdot\vecV_S\nonumber\\&&+\frac\theta2\int_{-\infty}^T\d T'\vecQ_{T\leftarrow T'}\cdot\bGamma(\vecR_{T'})\cdot\int_{-\infty}^{T'}\d S\vecB(\vecR_{T'},T';S)\cdot\sqrt{2\bGamma^{-1}(\vecR_S)}\cdot\bXi_S.\label{eq:RT_BBO_Vimplicit}
\end{eqnarray}

The advantage of the implicit Eq.~\eqref{eq:RT_BBO_Vimplicit} is that to order $\epsilon$, 
\begin{equation}
    \vecQ_{T\leftarrow T'}\to\bGamma^{-1}(\vecR_T)\delta(T-T').\label{eq:Qtransition}
\end{equation}
However, we must take this limit with caution: The autocorrelation of the fluctuating second term in Eq.~\eqref{eq:RT_BBO_Vimplicit} is of order $\epsilon^{-1}$, so the second term is of order $\epsilon^{-1/2}$. Therefore, terms such as this must be expanded to higher order, which we perform in Appendix~\ref{appendix:overdamped}. It is well known and we confirm in Appendix~\ref{appendix:overdamped} that the $\theta=0$ overdamped ($\epsilon\to0$) Langevin equation becomes~\cite{LauPRE2007}
\begin{equation}
    \vecV_T=\bGamma^{-1}(\vecR_T)\cdot \vecF(\vecR_T,T)+\grad\cdot\bGamma^{-1}(\vecR_T)+\sqrt{2\bGamma^{-1}(\vecR_T)}\cdot\bXi_T,\label{eq:overdamped_spurious}
\end{equation}
where the second term is often called the ``spurious drift'' and the third term is involves an It\^o product. The combination of a spurious drift and an It\^o product ensure that Boltzmann statistics are recovered under a conservative force~\cite{LauPRE2007}.

Upon including the remaining order-$\theta$ terms in Eq.~\eqref{eq:RT_BBO_Vimplicit}, as we explain in Appendix~\ref{appendix:overdamped}, the last order-$\theta$ noise term of Eq.~\eqref{eq:RT_BBO_Vimplicit} does not contribute to a spurious drift. Thus, after a single insertion iteration of Eq.~\eqref{eq:overdamped_spurious} in Eq.~\eqref{eq:RT_BBO_Vimplicit}, we obtain the overdamped Langevin equation corresponding to Eq.~\eqref{eq:RT-Langevin},
\begin{eqnarray}
    \frac{\d\vecR_T}{\d T}&=&\bGamma^{-1}(\vecR_T)\cdot\vecF(\vecR_T,T)-\theta\int_{-\infty}^{T}\d S\vecB(\vecR_T,T;S)\cdot\bGamma^{-1}(\vecR_S)\cdot\vecF(\vecR_S,S)\nonumber\\&&+\grad\cdot\bGamma^{-1}(\vecR_T)-\theta\int_{-\infty}^{T}\d S\vecB(\vecR_T,T;S)\cdot[\grad\cdot\bGamma^{-1}(\vecR_S)]\nonumber\\&&+\sqrt{2\bGamma^{-1}(\vecR_T)}\cdot\bXi_T-\frac\theta2\int_{-\infty}^{T}\d S\vecB(\vecR_T,T;S)\cdot\sqrt{2\bGamma^{-1}(\vecR_S)}\cdot\bXi_S.\label{eq:RT_overdamped_final}
\end{eqnarray}
Upon reintroducing dimensions, we obtain
\begin{eqnarray}
    \frac{\d\vecr_t}{\d t}&=&\bgamma^{-1}(\vecr_t)\cdot\vecf(\vecr_t,t)-\rho_\f\int_{-\infty}^{t}\d s\vecb(\vecr_t,t;s)\cdot\bgamma^{-1}(\vecr_s)\cdot\vecf(\vecr_s,s)\nonumber\\&&+\kt\grad\cdot\bgamma^{-1}(\vecr_t)-\kt\rho_\f\int_{-\infty}^{t}\d s\vecb(\vecr_t,t;s)\cdot[\grad\cdot\bgamma^{-1}(\vecr_s)]\nonumber\\&&+\sqrt{2\kt\bgamma^{-1}(\vecr_t)}\cdot\bxi_t-\frac{\rho_\f}2\int_{-\infty}^{t}\d s\vecb(\vecr_t,t;s)\cdot\sqrt{2\kt\bgamma^{-1}(\vecr_s)}\cdot\bxi_s.\label{eq:RT_overdamped_unit}
\end{eqnarray}
\end{widetext}
Equation~\eqref{eq:RT_overdamped_unit} is the final form of our second result in this work. Once the hydrodynamic kernel $\vecb$ is found for the corresponding problem geometry and fluid flow, Eq.~\eqref{eq:RT_overdamped_unit} contains all late-time memory effects without needlessly carrying over the colloids negligible inertia during the exploration times of $\vecf$. Expectedly from a problem with a position-dependent diffusivity $\kt\bgamma^{-1}$, Eq.~\eqref{eq:RT_overdamped_unit} contains a spurious drift which, interestingly, is ``memorized'' by the hydrodynamic kernel as well (see fourth term on the RHS). Indeed, upon specifying to spherical colloid in an unbounded fluid, $\bgamma=6\pi\eta a\,\vecI$ and $\vecb(t;s)=\vecI a(\pi\rho_\f\eta)^{-1/2}(t-s)^{-1/2}(\d/\d s)$ (where we dropped the added-mass inertia term), Eq.~\eqref{eq:RT_overdamped_unit} reduces to Eq.~\eqref{eq:OBLE}.

\section{Discussion}\label{sec:discussion}

In this work, we derived in closed form the Langevin equation of a colloidal particle within an arbitrary flow, confinement, and forcing. Using the RT, we obtain the full equation of motion of the colloid, $\d\vecv_t/\d t$, generalizing the FBBOE to arbitrarily shaped particles subject to arbitrary hydrodynamic memory. Taking note that the inertia of colloids is often negligible (quantified by a small inverse Schmidt number, $\mathrm{Sc}^{-1}\ll1$), through nondimensionalization of the full RT Langevin equation we found that colloid inertia can be subdominant compared to hydrodynamic memory. This implies that colloid inertia can be eliminated all the while hydrodynamic memory carries over to the overdamped (colloid-inertia-less) limit. Through stochastic Taylor expansions, we derived the overdamped RT Langevin equation, $\d\vecr_t/\d t$, thereby allowing the treatise of colloids subject to hydrodynamic memory without needlessly carrying its inertia. With that, we provide a formalism to study both underdamped and overdamped particle dynamics given knowledge of the hydrodynamics of the host liquid under a specified confinement. In our companion Letter~\cite{LETTER}, we studied the consequences of subjecting a trapped colloidal particle to power-law hydrodynamic memory; in this work, therefore, we provide the basis for extending that treatise to geometries beyond a spherical particle in an unbounded fluid.

In this work, we ignored fluid compressibility and considered a low Reynolds number. First, as a typical fluid's speed of sound $c$ far exceeds the typical velocities of trapped colloids, the fluid's density is able to exponentially equilibrate within $a/c$~\cite{ZwanzingJFM1975}, a timescale much shorter than the momentum diffusion time $\sim \rho_\f a^2/\nu$ ($\sim10^{-9}\,\mathrm{s}$ versus $\sim10^{-6}\,\mathrm{s}$, respectively, for a micron-sized colloid in water). Second, given the typical particle velocity $(\kt/w)/(\eta a)$ and traversed distance $w$, the Reynolds number coincides with the inverse-Schmidt number, $\mathrm{Re}=\rho_\f\kt/\eta^2 a=\mathrm{Sc}^{-1}\sim\epsilon$. Upon repeating the RT derivation in Sec.~\ref{sec:underdamped} for a flow with nonzero Reynolds number, one may confirm that the only modification in Eq.~\eqref{eq:RT-dimensional_almost} would have been $\partial\vecu/\partial t\to\partial\vecu/\partial t+(\vecu\cdot\grad)\vecu$ in the second term. As that memory term is already of order $\theta\sim\epsilon^{1/2}$ in our nondimensionalization, the nonlinear fluid-inertial term would have been of order $\theta\times\mathrm{Re}\sim\epsilon^{3/2}$, which is therefore truly negligible. Therefore, it is justified to start from Eq.~\eqref{eq:fluctuating-stokes}, involving an incompressible Stokes flow.

The agnosticism of our two central results\,---\,the underdamped and overdamped RT Langevin equations, Eqs.~\eqref{eq:RT-dimensional} and~\eqref{eq:RT_overdamped_unit}\,---\,to system details comes at the cost that one must supply the geometry's Green's function and full flow profiles to compute the  friction $\bgamma$ and kernel $\vecb$. These may require involved methods to compute in complex geometries~\cite{KimBOOK2013,StakgoldBOOK1979}. At first, this work seems to be mostly useful to describe particle dynamics for which the hydrodynamics problem is understood. However, we envision that Eqs.~\eqref{eq:RT-dimensional} and~\eqref{eq:RT_overdamped_unit} can be also used as a means of inference, as they provide velocity- and position-related observables, respectively. Using the counterpart of Eqs.~\eqref{eq:RT-dimensional} and~\eqref{eq:RT_overdamped_unit} to a spherical particle in an unbounded fluid, we show in our companion Letter~\cite{LETTER} that the $\sim t^{-3/2}$ scaling of the memory kernel manifests through $\sim t^{-3/2}$ power-law relaxation in autocorrelation functions. Thus, if considering hard-to-solve confinements, by resolving the late-time power-law relaxation, one may infer the hydrodynamic kernels. For instance, in a semi-infinite box, the dominant power is still $t^{-3/2}$~\cite{FelderhofJPCB2005}, whereas in a slab geometry the anisotropic kernel appears to reduce to $t^{-5/2}$ and $t^{-7/2}$, depending on orientation~\cite{FranoschPRE2009,JeneyPRL2008}. It would furthermore be interesting to devise systems where the memory kernel is not a power law, \eg amid momentum-absorbing scatterers, which may reduce the memory to exponential. Finally, our work is not limited to just passive particles, as activity may enter through the hydrodynamics kernels and force~\cite{TurkJFM2025}. Thus, in addition to passive microfluidic chambers, we envision applications to studying the rheology of confined active swimmers~\cite{ZhangNP2021} or bacteria subject to surface effects~\cite{BeerME2019}.

\begin{acknowledgements}

B.S. was supported by the Princeton Center for Theoretical Science and in part by the Center for the Physics of Biological Function at Princeton University. H.A.S and G.T. acknowledge support from the U.S. National Science Foundation via grant
No. CBET-2246791 and through the Princeton Center for Complex Materials (DMR-2011750).

\end{acknowledgements}

\appendix

\section{Fluctuating Basset–Boussinesq–Oseen equation from the reciprocal theorem}\label{appendix:BBO}

Here, we demonstrate how the general Eq.~\eqref{eq:RT-dimensional} specializes to Eq.~\eqref{eq:FBBOE} for a no-slip sphere of radius $a$, particle density $\rho_\p$, and velocity $\vecv_t$ in an unbounded fluid of density $\rho_\f$ and viscosity $\eta$. In this system, translation and rotation are decoupled, and for brevity, we ignore rotation in this appendix. The steady Stokes resistance is $\bgamma=6\pi\eta a\,\vecI$ and the translational RT equation is
\begin{align}
    \frac43\pi a^3\rho_\p\dot{\vecv}_t
    =&
    -6\pi\eta a\vecv_t
    -6\pi\eta a\rho_\f
    \int \veck(\vecr_t,\vecr)\cdot\frac{\partial\vecu(\vecr,t)}{\partial t}\d \vecr
    \nonumber\\&+\vecf(\vecr_t,t)+\sqrt{2\kt(6\pi\eta a)}\,\bta_t,
    \label{eq:BBO-RT-sphere}
\end{align}
with
\begin{equation}
    \veck(\vecr,\vecr')=
    \left(1+\frac{a^2}{6}\nabla'^2\right)\vecg(\vecr,\vecr')
    \label{eq:BBO-K}
\end{equation}
for a spherical particle. In an unbounded fluid, 
\begin{equation}
    \mathbf{G}(\vecr,\vecr')=\frac1{8\pi\eta|\vecr-\vecr'|}\left[\vecI+\frac{(\vecr-\vecr')(\vecr-\vecr')} {|\vecr-\vecr'|^2}\right].
\end{equation}

The only term we should simplify in Eq.~\eqref{eq:BBO-RT-sphere} is the RT volume integral (the third term), which we will simplify in Laplace space. We first note that both the kernel $\veck$ and velocity $\vecu$ are centered around the evolving particle position, preventing a convenient factorization of the convolution via a Laplace transform. To proceed, we will use the small-Reynolds-number condition underlying the BBO equation as follows: Both $\veck$ and $\vecu$ vary in space on the scale of particle size, $a$. The vorticity-diffusion time, over which they are built, is $a^2/(\eta/\rho_\f)$, during which the particle position $\vecr_t$ advances as far as $v\times \rho_\f a^2/\eta$, where $|\vecv(s)|\sim v$ is the typical particle velocity. The ratio between the particle displacement, $\rho_\f va^2/\eta$, and the range of variation of $\veck$ and $\vecu$, $a$, coincides with the Reynolds number $\mathrm{Re}=\rho_\f va/\eta$, which is assumed small throughout. Therefore, we may proceed to solve the RT volume integral Eq.~\eqref{eq:BBO-RT-sphere} where $\vecr_t$ is a frozen configuration (a constant parameter). 

We change variables $\vecr'\to\vecr-\vecr_t$ (where $\mathrm{Re}\ll1$ implies that $\d\vecr'/\d t\simeq\mathbf{0}$), and Laplace transform the convolution (second term) in Eq.~\eqref{eq:RT-dimensional} as follows
\begin{equation}
    6\pi\eta a\rho_\f\int_{|\vecr|>a}\veck(\mathbf{0},\vecr)\cdot\left[s\tilde{\vecu}(\vecr,s)\right]\d V
    =
    \Lambda(s)\tilde{\vecv}(s),
    \label{eq:BBO-Lambda-def}
\end{equation}
where rotational symmetry leaves only the direction of $\tilde{\vecv}(s)$. We proceed to evaluate the response coefficient $\Lambda(s)$.


In Laplace space, the unsteady Stokes response can be split into a motion-driven part, a bulk-noise-driven part, and initial-condition-determined part. We suppose that a particle has been diffusing within the liquid for a while, so we will ignore the last contribution. Furthermore, since the fluctuation-dissipation theorem is satisfied (Eq.~\eqref{eq:fluid-FDT}), the contribution to noise can be related to the friction kernel once all deterministic terms are found via Eq.~\eqref{eq:RT-BBO_noise}. Thus, in this section, we only need to consider the deterministic contribution, driven by the colloid motion. In Laplace space, the (noiseless) Stokes equation (Eq.~\eqref{eq:fluctuating-stokes}) is
\begin{align}
    \grad\cdot\tilde{\vecu}(\vecr,s)=&0,\nonumber\\
    \eta\left(\nabla^2-\frac{\rho_\f s}\eta\right)\tilde{\vecu}(\vecr,s)-\grad\tilde p(\vecr,s)=&0,
    \label{eq:BBO-unsteady-stokes}
\end{align}
with the boundary conditions $\tilde{\vecu}(|\vecr|=a,s)=\tilde{\vecv}(s)$ on the sphere and $\tilde{\vecu}(|\vecr|\to\infty,s)=0$ at infinity. 

Introducing the axisymmetric streamfunction, $\tilde\psi(r,\vartheta,s)=-|\vecv(s)|\sin^2\vartheta \phi(r,s)$, the velocity field takes the form
\begin{align}
    \tilde u_r(r,\vartheta,s)=&-\frac{2|\tilde\vecv(s)|\cos\vartheta}{r^2}\phi(r,s),
    \nonumber\\
    \tilde u_\vartheta(r,\vartheta,s)=&\frac{|\tilde\vecv(s)|\sin\vartheta}{r}\frac{\partial \phi(r,s)}{\partial r}.
    \label{eq:BBO-velocity-components}
\end{align}
Using Eq.~\eqref{eq:BBO-unsteady-stokes}, $\phi(r,s)$ satisfies
\begin{equation}
    \left(\frac{\d^2}{\d r^2}-\frac{2}{r^2}\right)\left(\frac{\d^2}{\d r^2}-\frac{2}{r^2}-\frac{\rho_\f s}\eta\right)\phi(r,s)=0.
    \label{eq:BBO-radial-ode}
\end{equation}
The decaying solution satisfying no slip on $r=a$ is
\begin{widetext}
\begin{equation}
    \phi(r,s)=
    -\frac{\eta a}{2\rho_\f s }\frac1r\left(\frac{\rho_\f s}\eta a^2+3\sqrt{\frac{\rho_\f s}\eta} a+3\right)
    +\frac{3a}{2}\sqrt{\frac\eta{\rho_\f s}}\exp\left[-\sqrt{\frac{\rho_\f s}\eta}(r-a)\right]\left(1+\sqrt{\frac\eta{\rho_\f s}}\frac1r\right).
    \label{eq:BBO-f-solution}
\end{equation}
Equipped with Eq.~\eqref{eq:BBO-f-solution}, we find
\begin{equation}
    \tilde{\vecu}(\vecr,s)
    =
    -\frac 1r\frac{\partial \phi(r,s)}{\partial r}\tilde{\vecv}(s)+\left[-\frac{2\phi(r,s)}{r^2}+\frac 1r\frac{\partial \phi(r,s)}{\partial r}\right]\frac{\vecr\vecr} {r^2}\cdot\tilde{\vecv}(s).
    \label{eq:BBO-motion-field}
\end{equation}
At the fixed particle center, the RT kernel can be written as
\begin{equation}
    \veck(\mathbf0,\vecr)=\frac{1}{8\pi\eta r}
    \left[
    \left(1+\frac{a^2}{3r^2}\right)\vecI+\left(1-\frac{a^2}{r^2}\right)\frac{\vecr\vecr}{r^2}
    \right].
    \label{eq:BBO-K-radial}
\end{equation}
Using $\int (\vecr\vecr/r^2)\sin\vartheta\d \vartheta\d\varphi=(4\pi/3)\vecI$, Eq.~\eqref{eq:BBO-Lambda-def} reduces to the radial integral
\begin{equation}
    \Lambda(s)
    =
    2\pi a\rho_\f s
    \int_a^\infty
    r\left\{
    -\frac 2r\frac{\partial \phi(r,s)}{\partial r}
    +\left(1-\frac{a^2}{3r^2}\right)\left[-\frac{2\phi(r,s)}{r^2}+\frac 1r\frac{\partial \phi(r,s)}{\partial r}\right]
    \right\}\d r.
    \label{eq:BBO-Lambda-radial}
\end{equation}
Substituting Eq.~\eqref{eq:BBO-f-solution} and defining $R=r/a$ and $S^{1/2}=(\rho_\f s/\eta)^{1/2} a$ (for they are the nondimensional $r$ and $s$, respectively) gives
\begin{equation}
    \Lambda\left(\frac{\eta S}{\rho_\f a^2}\right)
    =
    \pi\eta a
    \int_1^\infty 
    \left[(S+3S^{1/2}+3)\left(\frac1{R^2}-\frac1{R^4}\right)\right.
    \left.+3e^{-S^{1/2}(R-1)}
    \left(
    S-\frac{S^{1/2}}{R}-\frac{1-S/3}{R^2}+\frac{S^{1/2}}{R^3}+\frac1{R^4}
    \right)
    \right]\d R.
    \label{eq:BBO-Lambda-I}
\end{equation}
\end{widetext}
Upon solving the integral, we find
\begin{equation}
    \Lambda(s)=\frac23\pi a^3\rho_\f s+6\pi a^2\sqrt{\rho_\f \eta s}.
    \label{eq:BBO-Lambda}
\end{equation}
The first term is the added-mass contribution and the second is the Laplace-space Basset kernel. Thus, inserting Eq.~\eqref{eq:BBO-Lambda} into~\eqref{eq:BBO-Lambda-def}, and inverting the Laplace transform, we recover Eq.~\eqref{eq:FBBOE}, where the noise is enforced to satisfy the fluctuation-dissipation theorem, Eq.~\eqref{eq:BBO_noise}.

\section{Scaling of the hydrodynamic-memory term}\label{appendix:sqrt-epsilon}

We discuss the generality of the scaling argument behind $\theta=\epsilon^{1/2}$ prefactor in Eq.~\eqref{eq:RT-Langevin}. At the core of the argument lies the fact that in Laplace space, the unsteady Stokes operator introduces the Basset length $(\rho_\f s/\eta)^{-1/2}$ in an unbounded fluid. This length is the distance over which a time-dependent disturbance penetrates into the fluid.

For a translation of a sphere, the motion-driven flow separates into a steady algebraic part and an exponentially screened unsteady part,
\begin{gather}
    \tilde{\vecu}
    =
    \tilde{\vecu}_{\mathrm{st}}
    +
    \tilde{\vecu}_{\mathrm{un}},
    \quad
    \tilde{\vecu}_{\mathrm{st}}\sim r^{-3}|\tilde\vecv(s)|,
    \nonumber\\
    \tilde{\vecu}_{\mathrm{un}}\sim \frac{a}{r}e^{-(\rho_\f s/\eta)^{1/2}(r-a)}.
    \label{eq:translation-flow-split}
\end{gather}
These two pieces contribute separately to the RT volume integral in Eq.~\eqref{eq:BBO-Lambda-def}. The algebraic part gives the added-mass contribution, proportional to $s$, while the screened part gives the Basset memory. For the latter, angular integration only affects the prefactor. With $\veck\sim1/(\eta r)$, unsteady velocity amplitude $\tilde{\vecu}_{\mathrm{un}}$, and volume element $r^2\d r$, we find the scalings
\begin{equation}
    \Lambda(s)\sim \eta a\rho_\f s\int_a^\infty\frac{1}{\eta r}\frac{ae^{-(\rho_\f s/\eta)^{1/2}(r-a)}}{r}r^2\d r=a^2\sqrt{s \rho_\f\eta}.
    \label{eq:translation-scaling}
\end{equation}
The factor $s$ from $\partial/\partial t$ is therefore partially canceled by the cutoff Basset length $(\rho_\f s/\eta)^{-1/2}$, leaving a kernel that scales as $s^{1/2}$. Normalizing the kernel by the steady Stokes drag, $\Lambda(s)/(6\pi\eta a)\sim (s \rho_\f/\eta)^{1/2}a$, and nondimensionalizing time $S=w^2 s/D$, we find precisely $\Lambda(s)/(6\pi\eta a)\sim (\epsilon S)^{1/2}$, where $\epsilon$ is that of Eq.~\eqref{eq:epsilon} (up to numerical prefactors).

The same power occurs for rotation. The steady rotational kernel decays faster, $\veck\sim a^3/(\eta r^2)$, whereas the exact unsteady rotational field carries the relevant $s$ dependence. For a sphere rotating with angular velocity $\tilde{\boldsymbol{\nu}}(s)$, Lamb's solution gives the azimuthal velocity~\cite{LambBOOK1932}
\begin{align}
    \tilde u_\varphi(r,\vartheta,s)
    =&
    |\tilde{\boldsymbol{\nu}}(s)| a\sin\vartheta
    \nonumber\\&\times\frac{a^2}{r^2}\frac{1+(\rho_\f s/\eta)^{1/2} r}{1+(\rho_\f s/\eta)^{1/2} a}
    e^{-(\rho_\f s/\eta)^{1/2}(r-a)}.
    \label{eq:rotation-flow}
\end{align}
Unlike the translational field, this is not naturally a sum of a separate algebraic Stokes part and a screened unsteady correction. Instead, the full field is screened and the normalization $[1+(\rho_\f s/\eta)^{1/2} a]^{-1}$ carries the long-time expansion. After angular integration, the radial dependence entering the RT volume integral contains the integral $\int_a^\infty [1+(\rho_\f s/\eta)^{1/2} r] e^{-(\rho_\f s/\eta)^{1/2}(r-a)}/r^2\d r=1/a$. Thus, the $s$ dependence comes from the late-time expansion $[1+(\rho_\f s/\eta)^{1/2} a]^{-1}\simeq1-(\rho_\f s/\eta)^{1/2} a$. The leading-order term gives rise to the steady-Stokes contribution, and the second to Basset memory term. By taking the ratio among the two, we once again we identify $\Lambda(s)/(8\pi\eta a^3)\sim (s \rho_\f/\eta)^{1/2}a=(\epsilon S)^{1/2}$.

For an unbounded sphere, the above arguments were made exact to give the classical BBO prefactors (see Appendix~\ref{appendix:BBO}). For a nonspherical particle, the near-field unsteady Stokes response still contains a Basset layer and the $s^{1/2}$ power is robust, with geometry entering the prefactor and tensor structure. Confinement can modify the kernel~\cite{FranoschPRE2009,JeneyPRL2008}. At the diffusive time $w^2/[\kt/(\eta a)]$, the Basset length is $\epsilon^{-1/2}a$ if \(w\sim a\), so nearby walls lie within the vorticity-diffusion layer. For a sphere near an infinite planar wall, interestingly, Felderhof's leading-order result in $a/h$ ($h$ is the distance to the wall) leaves the Basset coefficient unchanged; the leading dynamical wall correction enters at order $s^{3/2}$, which nondimensionalizes as $\mathcal{O}(\epsilon^{3/2})$, subleading to the $s^{1/2}$ term~\cite{FelderhofJPCB2005}. More general confinement, such as a finite enclosure~\cite{FranoschPRE2009,JeneyPRL2008}, reduce the long-time functional form. In the RT framework, these changes are carried by the geometry-dependent kernel $\mathsf{k}(\vecR,\vecR')$; so long as the corrections from hydrodynamic effects can be argued to be small ($\theta\ll1$), the procedure described in the text should still hold, and the emergent power-law memories would asymptotically dominate over exponential Stokesian relaxation.

\section{Stochastic expansions to eliminate colloid inertia}\label{appendix:overdamped}

In this appendix, we fill in the technical details in deriving Eq.~\eqref{eq:RT_overdamped_final} starting from Eq.~\eqref{eq:RT_BBO_Vimplicit}. First, it is instructive to show how the spurious drift $\grad\cdot\bGamma^{-1}(\vecR_T)$ arises in Eq.~\eqref{eq:overdamped_spurious}, starting from a free particle without hydorydnamic effects\,---\,Eq.~\eqref{eq:RT_BBO_Vimplicit} with $\theta=0$ and the absence of force ($\vecf=\mathbf0$),
\begin{equation}
    \vecV_T
    =\int_{-\infty}^{T}\d T'\vecQ_{T\leftarrow T'}\cdot\sqrt{2\bGamma(\vecR_{T'})}\cdot\bXi_{T'}.\label{eq:underdamped-Vimplicit}
\end{equation}
A common technique is to arrive at the spurious drift via the corresponding Fokker-Planck equation~\cite{SchussBOOK2010,GardinerBOOK2009}. However, the Fokker-Planck equation corresponding to Eq.~\eqref{eq:FBBOE} contains memory~\cite{ChowJCP1972}, complicating the common procedure when we will aim to derive \eqref{eq:RT_overdamped_final}. 

We instead generalize the conventional method to eliminate colloid inertia~\cite{SchussBOOK2010,GardinerBOOK2009} to the case of a position-dependent friction tensor $\bGamma(\vecR)$. Note that $\langle\vecV_T\vecV_T\rangle\sim1/\epsilon$ according to Eq.~\eqref{eq:underdamped-Vimplicit}, meaning that the velocity's correlation time shrinks as $\epsilon$ but $\vecV_{T'}$'s amplitude at $T'\to T$ diverges as $1/\epsilon^{1/2}$. Since $\vecQ_{T\leftarrow S}$ depends on the history $\{\vecR_S\}_{S\in[T,T')}$ through $\bGamma(\vecR_T)$, which changes over time as $\vecR_T-\vecR_{T'}\sim\epsilon^{1/2}$, a Taylor expansion of $\vecR_{T'}$ around $\vecR_T$ will contribute an order-$1$ term overall to $\vecV_T$. Thus, to take the limit of $\epsilon\to0$ and use Eq.~\eqref{eq:Qtransition}, we must expand $\vecQ_{T\leftarrow T'}$ to a higher order. Indeed, we expect a deterministic, so-called spurious drift term to appear in the expanded Eq.~\eqref{eq:underdamped-Vimplicit}~\cite{LauPRE2007}, 
\begin{equation}
    \vecV_T
    =\grad\cdot\bGamma^{-1}(\vecR_T)+\sqrt{2\bGamma^{-1}(\vecR_T)}\cdot\bXi_{T'}.\label{eq:overdamped_free_posi}
\end{equation}

To derive the ovrerdamped Eq.~\eqref{eq:overdamped_free_posi}, we write $\vecR_{T''}=\vecR_{T'}+\bDelta_{T''\leftarrow T'}$ where, using Eq.~\eqref{eq:underdamped-Vimplicit}, the increment $\bDelta_{T''\leftarrow T'}=\int_{T'}^{T''}\vecV_S\d S$ satisfies
\begin{equation}
    \bDelta_{T''\leftarrow T'}=\int_{T'}^{T''}\d S\int_{-\infty}^S\d S'\vecQ_{S\leftarrow S'}\cdot\sqrt{2\bGamma(\vecR_{S'})}\cdot\bXi_{S'}.\label{eq:DeltaIncrement}
\end{equation}
With $T'<T''<T$ in Eq.~\eqref{eq:underdamped-Vimplicit}, note that only $T-T'=\calO(\epsilon)$ contribute (as the contribution from earlier $T'$s are exponentially suppressed). Therefore, in both Eqs.~\eqref{eq:DeltaIncrement} and~\eqref{eq:underdamped-Vimplicit}, $T',T'',S,S',S''=T-\calO(\epsilon)$ and thus $\bDelta_{T''\leftarrow T'}=\calO(\epsilon^{1/2})$. This allows us to expand in small $\bDelta_{T''\leftarrow T'}$, \eg $\bGamma(\vecR_{T''})=\bGamma(\vecR_{T'})+(\bDelta_{T''\leftarrow T'}\cdot\grad)\bGamma(\vecR_{T'})+\calO(\epsilon)$, with which we write the first-order perturbation to a matrix exponential~\cite{ZwanzigBOOK2001}
\begin{widetext}
\begin{equation}
    \vecQ_{T\leftarrow T'}=\vecQ_{T\leftarrow T'}^0+\int_{T'}^T\d T''\vecQ_{T\leftarrow T''}^0\cdot(\bDelta_{T''\leftarrow T'}\cdot\grad)\bGamma(\vecR_{T'})\cdot\vecQ_{T''\leftarrow T'}^0+\calO(\epsilon),\label{eq:Q-expand}
\end{equation}
where
\begin{equation}
    \vecQ_{T\leftarrow T'}^0=e^{-(\epsilon\vecM)^{-1}\cdot\bGamma (\vecR_{T'})(T-T')}\cdot(\epsilon\vecM)^{-1}.
\end{equation}
Note that we deliberately expanded around $T'$, so ultimately the It\^o product would emerge. 

Equation~\eqref{eq:Q-expand} should now be inserted in Eq.~\eqref{eq:underdamped-Vimplicit},
\begin{eqnarray}
    \vecV_T
    &=&\int_{-\infty}^{T}\d T'\vecQ_{T\leftarrow T'}^0\cdot\sqrt{2\bGamma(\vecR_{T'})}\cdot\bXi_{T'}\nonumber\\&&-\int_{-\infty}^{T}\d T'\int_{T'}^T\d T''\vecQ_{T\leftarrow T''}^0\cdot(\bDelta_{T''\leftarrow T'}\cdot\grad)\bGamma(\vecR_{T'})\cdot\vecQ_{T''\leftarrow T'}^0\cdot\sqrt{2\bGamma(\vecR_{T'})}\cdot\bXi_{T'}.\label{eq:underdamped_Vimplicit_interrim}
\end{eqnarray}
In the first term of Eq.~\eqref{eq:underdamped_Vimplicit_interrim}, the integrand only dependents $T'$ (not the future), and hence is decorrelated from $\bXi_{T'}$. Thus, the first term's covariance between times $T,S$ is simply
\begin{equation}
    2\int_{-\infty}^{\min(T,S)}\d T'\vecQ_{T\leftarrow T'}^0\cdot\bGamma(\vecR_{T'})\cdot\vecQ_{S\leftarrow T'}^0\to 2\bGamma^{-1}(\vecR_{T})\delta(T-S),
\end{equation}
which thus may be expressed in terms of a white noise, Eq.~\eqref{eq:white_noise_nondim_FT}. The second term of Eq.~\eqref{eq:underdamped_Vimplicit_interrim} is already of order $1$, and hence its fluctuations will be of higher order in $\epsilon$. In total, we may replace the outer product $\bDelta_{T''\leftarrow T'}\bXi_{T'}$ with its average,
\begin{equation}
    \langle\bDelta_{T''\leftarrow T'}\bXi_{T'}\rangle=\int_{T'}^{T''}\d S\vecQ_{S\leftarrow T'}\cdot\sqrt{2\bGamma(\vecR_{T'})}\label{eq:ave}
\end{equation}
where we computed the friction $\bGamma(\vecR_{T'})$ to leading order in $\epsilon$ as this is already a correction term. Thus, the second term along the $i$th axis reads
\begin{align}
    2&\int_{-\infty}^{T}\d T'\int_{T'}^T\d T''\int_{T'}^{T''}\d S(\vecQ_{T\leftarrow T''}^0)_{ij}\nabla_l\Gamma_{jk}(\vecR_{T'})[\vecQ_{T''\leftarrow T'}^0\cdot\bGamma(\vecR_{T'})\cdot\vecQ_{S\leftarrow T'}^{0\T}]_{kl}\nonumber\\&\to2[\bGamma^{-1}(\vecR_T)]_{ij}\nabla_l\Gamma_{jk}(\vecR_{T})[\bGamma^{-1}(\vecR_T)\cdot\bGamma(\vecR_T)\cdot \bGamma^{-1}(\vecR_T)]_{kl}\nonumber\\&\times\int_{-\infty}^{T}\d T'\int_{T'}^T\d T''\int_{T'}^{T''}\d S\delta(T-T'')\delta(T''-T')\delta(S-T')=-\grad\cdot\bGamma^{-1}(\vecR_T),
\end{align}
\end{widetext}
with summation over repeated spatial indices implied. Combined, we obtain Eq.~\eqref{eq:overdamped_free_posi}. 

The above clearly shows that we only encounter complications with using Eq.~\eqref{eq:Qtransition} if $\vecQ_{T\leftarrow T'}$ multiplies $\bXi_T$, in which case a careful stochastic Taylor expansion should be undertaken. This does mean, however, than deterministic integral involving $\vecQ_{T\leftarrow T'}$ (\ie first term in Eq.~\eqref{eq:RT_BBO_Vimplicit}) can immediately utilize Eq.~\eqref{eq:Qtransition}.

Since the hydrodynamic memory (appearing in the third and fourth terms in Eq.~\eqref{eq:RT_BBO_Vimplicit}) is an integral over white noise, ignoring the prefactor $\theta$, their covariance is of order $1$ in $\epsilon$ (as opposed to the covariance of the second term, singled out in Eq.~\eqref{eq:underdamped-Vimplicit}, whose covariance is order $1/\epsilon$). Thus, the corrections from higher-order stochastic Taylor expansion of $\vecQ_{T\leftarrow T'}$ are negligible. Therefore, they are subdominant in the above procedure, and should not be included in Eqs.~\eqref{eq:underdamped_Vimplicit_interrim} and~\eqref{eq:ave}. In conclusion, to obtain Eq.~\eqref{eq:RT_overdamped_final} from Eq.~\eqref{eq:RT_BBO_Vimplicit}, we use Eq.~\eqref{eq:Qtransition} in the first, third, and fourth term in Eq.~\eqref{eq:RT_BBO_Vimplicit}, replace the second term with Eq.~\eqref{eq:overdamped_free_posi}, and self-consistently replace $\vecV_S$ in the third term with the leading order expression in $\theta$, Eq.~\eqref{eq:overdamped_spurious}. With that, we derived the leading order inertia-eliminated ($\epsilon\to0$) hydrodynamically-corrected ($0<\theta\ll1$) overdamped Langevin equation.

\end{document}